\documentclass[reprint, amsmath,
]{revtex4-2}

\pdfoutput=1

\usepackage{multirow}
\usepackage{float}
\usepackage{graphicx}
\usepackage{caption}
\usepackage{dcolumn}
\usepackage{bm}

\begin{document}

\preprint{APS/123-QED}

\title{Machine Learning Percolation Model}

\author{Shu Cheng\textsuperscript{1}}
\author{Huai Zhang}
\email{hzhang@ucas.ac.cn}
\author{Yaolin Shi}
\affiliation{Key Laboratory of Computational Geodynamics, College of Earth and Planetary Sciences, University of Chinese Academy of Sciences, No.19(A) Yuquan Road, Shijingshan District, Beijing 100049, China}
\author{Fei He\textsuperscript{1}}
\author{Ka-Di Zhu}
\email{zhukadi@sjtu.edu.cn}
\affiliation{Key Laboratory of Artificial Structures and Quantum Control (Ministry of Education), School of Physics and Astronomy, Shanghai Jiao Tong University,
800 Dong Chuan Road, Shanghai 200240, China
}

\date{\today}

\begin{abstract}
Recent advances in machine learning have become increasingly popular in the applications of phase transitions and critical phenomena. By machine learning approaches, we try to identify the physical characteristics in the two-dimensional percolation model. To achieve this, we adopt Monte Carlo simulation to generate dataset at first, and then we employ several approaches to analyze the dataset. Four kinds of convolutional neural networks (CNNs), one variational autoencoder (VAE), one convolutional VAE (cVAE), one principal component analysis (PCA), and one $k$-means are used for identifying order parameter, the permeability, and the critical transition point. The former three kinds of CNNs can simulate the two order parameters and the permeability with high accuracy, and good extrapolating performance. The former two kinds of CNNs have high anti-noise ability. To validate the robustness of the former three kinds of CNNs, we also use the VAE and the cVAE to generate new percolating configurations to add perturbations into the raw configurations. We find that there is no difference by using the raw or the perturbed configurations to identify the physical characteristics, under the prerequisite of corresponding labels. In the case of lacking labels, we use unsupervised learning to detect the physical characteristics. The PCA, a classical unsupervised learning, performs well when identifying the permeability but fails to deduce order parameter. Hence, we apply the fourth kinds of CNNs with different preset thresholds, and identify a new order parameter and the critical transition point. Our findings indicate that the effectiveness of machine learning still needs to be evaluated in the applications of phase transitions and critical phenomena.
\end{abstract}


\maketitle


\section{\label{sec:Introduction} Introduction}

Machine learning methods have rapidly become pervasive instruments due to better fitting quality and predictive quality in comparison with traditional models in terms of phase transitions and critical phenomena. Usually, machine learning can be divided into supervised and unsupervised learning. In the former, the machine receives a set of inputs and labels. Supervised learning models are trained with high accuracy to predict labels. The effectiveness of supervised learning has been examined by many predecessors on Ising models \cite{tanaka2017detection,carrasquilla2017machine,van2017learning,shiina2020machine,suchsland2018parameter,huembeli2018identifying,Wetzel2017Machine,wetzel2017unsupervised}, Kitaev chain models \cite{van2017learning}, disordered quantum spin chain models \cite{van2017learning}, Bose-Hubbard models \cite{huembeli2018identifying}, SSH models \cite{huembeli2018identifying}, SU(2) lattice gauge theory \cite{Wetzel2017Machine}, topological states models \cite{deng2017machine}, $q$-state Potts models \cite{bachtis2020mapping,zhao2019machine}, uncorrelated configuration models \cite{ni2019machine}, Hubbard models \cite{ch2018unsupervised,ni2019machine}, and $XY$ models \cite{hu2017discovering}, ect. 

On the other hand, in unsupervised learning models, there are no labels. Unsupervised learning can be used as meaningful analysis tools, such as sample generation, feature extraction, cluster analysis. Principal component analysis (PCA) is one of the unsupervised learning techniques. Recently investigators have examined the PCA's effectiveness for exploring physical features without labels in the applications of phase transitions and critical phenomena \cite{Wang2016Discovering,hu2017discovering,wetzel2017unsupervised,ch2018unsupervised,Zhang2018Machine,yu2020unsupervised}. Variational autoencoder (VAE) and convolutional VAE (cVAE), another two classical unsupervised learning technique, incorporated into generative neural networks, are used for data reconstruction and dimensional reduction in respect of phase transitions and critical phenomena \cite{wetzel2017unsupervised,d2020learning}. 

Although machine learning approaches have been applied successfully in phase transitions and critical phenomena, there is only one study on the percolation model \cite{yu2020unsupervised}. Motivated by predecessors, we conduct much more comprehensive studies, which combine supervised learning with unsupervised learning, to detect the physical characteristics in the percolation model. 

Our work is considered from the following several aspects. First, we use the former three kinds of deep convolutional neural networks (CNNs) to deduce the two order parameters and the permeabilities in the two-dimensional percolation model. our inspiration and method come from \cite{carrasquilla2017machine,van2017learning}, whose study both focus on Ising model.

Nevertheless, the above CNNs are trained on the known configurations from the dataset obtained by Monte Carlo simulation. \cite{wetzel2017unsupervised,d2020learning} find that VAE and cVAE can reconstruct samples in Ising model. Hence, we use the VAE and the cVAE to generate new configurations that are out of the dataset. After generating the new configurations, we pour them into the former three kinds of CNNs, respectively. 

Having explored supervised learning, we now move on to unsupervised learning. Here we try to identify physical characteristics without labels in the two-dimensional percolation model. \cite{Wang2016Discovering} takes the first principal component obtained by PCA as the order parameter in Ising model. In contrast to \cite{Wang2016Discovering}, by using preprocessing on the unpercolating clusters, \cite{yu2020unsupervised} also successfully finds the order parameter in percolation model by PCA. In this study, we try to  use the PCA to extract relevant low-dimensional representations to discover physical characteristics.

In an actual situation, we may not know the labels when identifying order parameter. To overcome the difficulty associated with missing labels, \cite{ni2019machine} changes the preset threshold between the labels zero
and one so as to make incorrect labels between the preset and the true thresholds. Hence, we deliberately change the preset thresholds, determined by $k$-means, between the labels zero and one. Here we use the fourth kinds of CNNs which receives the raw configurations as input and the labels determined by the preset thresholds. 

This paper is organized as follows. In Sec.~\ref{sec:percolation}, we describe the two-dimensional percolation model and the dataset from Monte Carlo simulation. In Sec.~\ref{sec:machine_learning}, we give a brief introduction to CNNs, VAE and cVAE, and PCA. Next, we provide dozens of machine learning models to capture the physical characteristics and discuss the results in Sec.~\ref{sec:result_and_discussion}. Finally, we conclude with a summary in Sec.~\ref{sec:conclusion}.

\section{\label{sec:percolation} The two-dimensional Percolation model}

For percolation models, what we need to do is to capture the physical characteristics. A suitable dataset should be constructed to fulfill this objective. Various models in physical dynamics can be simulated mathematically by the Monte Carlo method, and it has been proved to be valid for using the Monte Carlo simulation to capture different physical features in phase transitions and critical phenomena \cite{Wang2016Discovering, hu2017discovering, wetzel2017unsupervised}. 

In this study, the Monte Carlo simulation for the two-dimensional percolation model is carried out as follows. First, 40 values of permeability range from 0.41 to 0.80 with an interval of 0.01. For each permeability, the initial samples consist of 1000 percolating configurations. To train the machine learning models, the matrix $\bm{X}$ with the size of $M\times N$ (see Eq.~\ref{eq:X}) is used for storing 40,000 raw percolating configurations.
\begin{eqnarray}
\bm{X} = \left(
\begin{array}{ccccc}
a_{1,1} & a_{1,2} & \ldots & a_{1,N-1} & a_{1,N}\\
a_{2,1} & a_{2,2} & \ldots & a_{2,N-1} & a_{2,N}\\
\vdots & \vdots & \ddots & \vdots & \vdots\\
a_{M-1,1} & a_{M-1,2} & \ldots & a_{M-1,N-1} & a_{M-1,N}\\
a_{M,1} & a_{M,2} & \ldots & a_{M,N-1} & a_{M,N}\\
\end{array} \right)_{M\times N} 
\label{eq:X}.
\end{eqnarray}

In Eq.~(\ref{eq:X}), $M=40,000$, $N=L\times L$, and $L=28$. $M$ and $N$ represent the number of the configurations and the lattices, respectively. Each row $\bm{R}_i$ ($i=1,2,\ldots,M$) in the matrix $\bm{X}$ is a configuration with one dimension, can be reshaped as the matrix $\bm{X}_i$ ($i=1,2,\ldots,M$) with the size of $L\times L$ (see Eq.~\ref{eq:x_i}). Furthermore, each column $\bm{C}_j$ ($j=1,2,\ldots,N$) in the matrix $\bm{X}$ represents one lattice with different configurations. Moreover, the element $a_{ij}$ ($i=1,2,\ldots,M;j=1,2,\ldots,N$) in matrix $\bm{X}$ and the element $b_{kl}$ ($k,l=1,2,\ldots,L$) in matrix $\bm{X}_i$
take 0 when the corresponding lattice is occupied and take 1 otherwise. 
\begin{eqnarray}
\bm{X}_i = \left(
\begin{array}{ccccc}
b_{11} & b_{12} & \ldots & b_{1 L-1} & b_{1L} \\
  &  & \vdots & & \\
b_{L1} & b_{L2} & \ldots & b_{L L-1} & b_{LL}\\
\end{array} \right)_{L\times L}
\label{eq:x_i}.
\end{eqnarray}

\begin{figure*}[htbp]
 \includegraphics[width=0.99\textwidth]{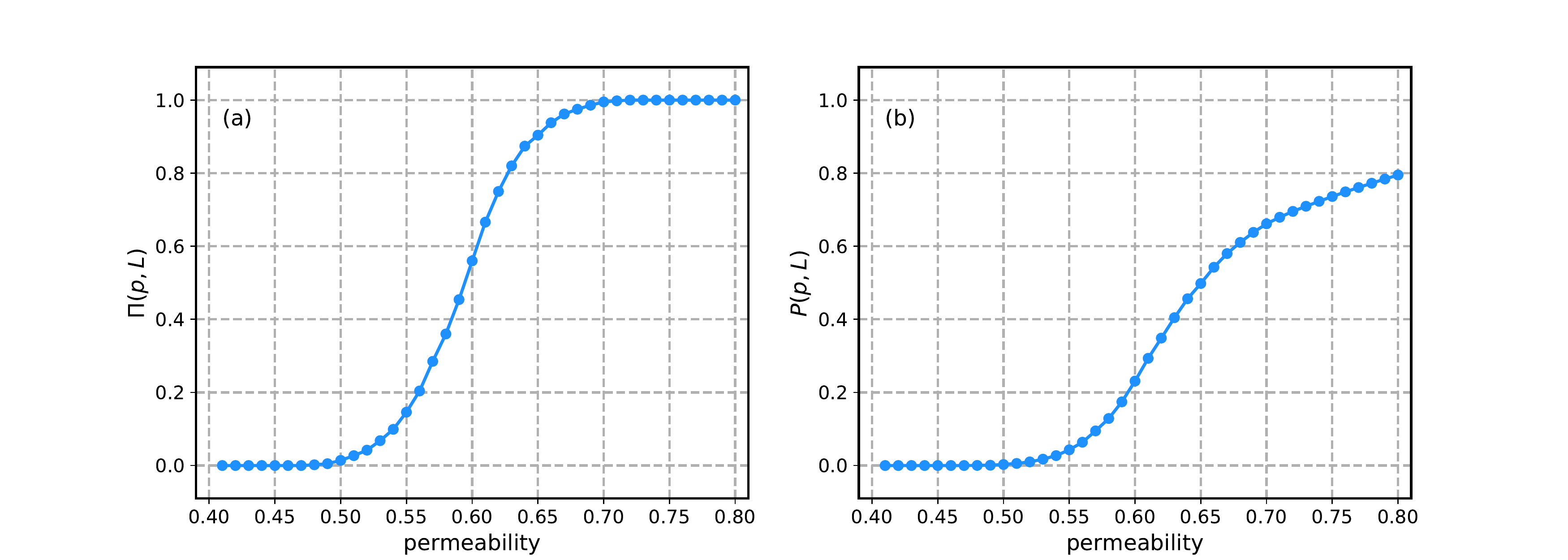}
 \caption{\label{fig:percolation-pi-p} (a) The relationship between the permeabilities $\{0.41,0.42,\ldots,0.80\}$ and the raw $\varPi(\bm{p},L)$. (b) The relationship between the permeabilities $\{0.41,0.42,\ldots,0.80\}$ and the raw $P(\bm{p},L)$.}
\end{figure*}

Except for the raw configurations, the dataset also encompasses order parameter. In the two-dimensional percolation model, order parameter includes the percolation probability $\varPi(\bm{p},L)$ and the density of the spanning cluster $P(\bm{p},L)$. $\varPi(\bm{p},L)$ refers to the probability that there is one connected path from one side to another in $\bm{X}_i$. That is to say, $\varPi(\bm{p},L)$ is a function of the permeability $\bm{p}$ in the system with the size of $L\times L$ (see Fig.~\ref{fig:percolation-pi-p}(a)). With a connectivity of 4, we can identify the cluster for each lattice $b_{kl}$. The clusters are marked sequentially with an unique index. Note that the two lattices having the same index belong to the same cluster. If there are more than one cluster, the greatest cluster is chosen as the result. In this way, we can count up how many times the configurations $\{\bm{X}_1,\bm{X}_2,\ldots,\bm{X}_M\}$ are percolated for given $\bm{p}$ and $L$. For each permeability $p$, $\varPi(\bm{p},L)_p$ is expressed in Eq.~(\ref{eq:pi}).
\begin{eqnarray}
    \varPi(\bm{p},L)_p=\frac{1}{1000}\sum_{r=1}^{1000}{S_r}
    \label{eq:pi},
\end{eqnarray}
Where $S_r$ refers that the two-dimensional configurations $\{\bm{X}_{e},\bm{X}_{2\times e},\ldots,\bm{X}_{1000\times e}\}$ for the permeability $p$ is percolated or not. Clearly, $S_r$ takes 0 if the corresponding lattice is occupied and takes 1 otherwise. $r=\{1,2,\ldots,1000\}$, $p\in \{0.41,0.42,\ldots,0.80\}$, $e=(p-0.40)\times 100$, and $\varPi(\bm{p},L)_p\in [0,1]$.

Another representation of order parameter is the density of the spanning cluster $P(\bm{p},L)$. In contrast to the $\varPi(\bm{p},L)$, $P(\bm{p},L)$ is associated with spanning cluster. Therefore, for all configurations $\{\bm{X}_1, \bm{X}_2, \ldots, \bm{X}_M\}$, $P(\bm{p},L)$ is characterized by that whether or not each lattice $b_{kl}$ belongs to the total spanning cluster. Similarly, $P(\bm{p},L)$ is a function of the permeability $\bm{p}$ in the system with the size of $L\times L$ (see Fig.~\ref{fig:percolation-pi-p}(b)). For each permeability $p$, $P(\bm{p},L)_p$ is expressed in Eq.~(\ref{eq:p}). 
\begin{eqnarray}
    P(\bm{p},L)_p=\frac{1}{1000\times L\times L}\sum_{r=1}^{1000}{S_r^{'}}
    \label{eq:p}.
\end{eqnarray}
Where $S_r^{'}$ counts up the total number of lattices that belong to the spanning cluster for each configuration in  $\{\bm{X}_{e},\bm{X}_{2\times e},\ldots,\bm{X}_{1000\times e}\}$ for the permeability $p$. Obviously, $r=\{1,2,\ldots,1000\}$, $p\in \{0.41,0.42,\ldots,0.80\}$, $e=(p-0.40)\times 100$, $0\le S_r^{'}<L\times L$, and $P(\bm{p},L)_p\in [0,1)$.

\section{\label{sec:machine_learning}Machine learning methods}

\subsection{\label{sec:cnn}CNNs}

In this section, we will focus on the two-dimensional percolation model and the dataset obtained by the Monte Carlo simulation. This section will discuss several machine learning approaches, including CNNs, VAE and cVAE, and PCA, to deduce physical characteristics.

Let us first introduce CNNs. CNNs, supervised learning methods, are particularly useful in solving realistic problem for many disciplines, such as physics\cite{van2017learning}, chemistry \cite{Janet2020Machine}, medicine \cite{Plant2020Machine}, economics \cite{Hull2021Machine}, biology \cite{Buchanan2020Machines}, and geophysics \cite{Poorvadevi2020An,cheng2020comparison}, ect. In the applications of phase transitions and critical phenomena, many predecessors utilize CNNs to detect physical features, especially order parameter \cite{tanaka2017detection,kashiwa2019phase,arai2018deep,xu2019new,bachtis2020mapping,suchsland2018parameter,carrasquilla2017machine,huembeli2018identifying}. In this study, the four kinds of CNNs are not only used to detect the two order parameters ($\varPi(\bm{p},L)$ and $P(\bm{p},L)$), but also to detect the permeability $\bm{p}$.

\begin{figure*}[htbp]
    \includegraphics[width=0.6\textwidth]{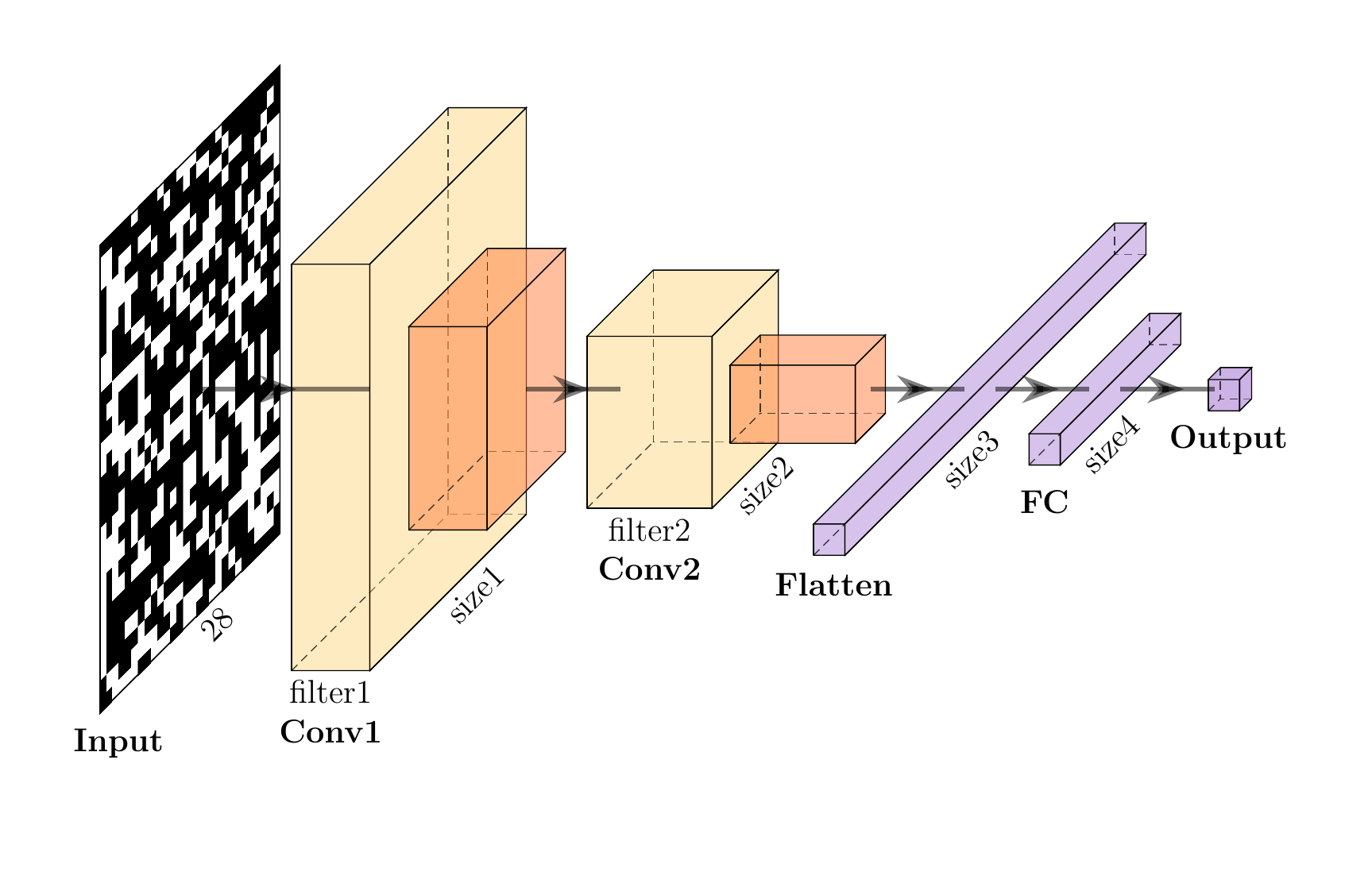}
    \caption{\label{fig:cnn} The structure of the CNNs with four layers, including ``Conv1'', ``Conv2'', ``FC'', and ``Output''. The square with black and white lattices is percolating configurations $\{\bm{X}_1, \bm{X}_2, \ldots, \bm{x}_M\}$. The light yellow cuboids (``Conv1'' and ``Conv2'') stand for convolution layers. The bright orange cuboids stand for max-pooling layers. The light purple cuboid ``FC'' refer to a fully connected layer. The input layer ``Input'' has the size of $28\times 28$. The first layer ``Conv1'' with ``filter1'' filters has the size of ``size1''$\times$``size1''. The second layer ``Conv2'' with ``filter2'' filters has the size of ``size2''$\times$``size2''. The layer ``Flatten'' owns the size of ``size3''. The third layer ``FC'' owns the size of ``size4''. And the last layer ``Output'' represents for a fully connected layer with one neuron.}
\end{figure*}

Next, we demonstrate the architecture of the CNNs (see Fig.~\ref{fig:cnn}). The structure of the CNNs has four layers, including two convolution layers and two fully connected layers. The percolating configurations $\{\bm{X}_1,\bm{X}_2,\ldots,\bm{X}_M\}$ are taken as inputs. Consequently, the CNNs receive the corresponding order parameters ($\varPi(\bm{p},L)$ and $P(\bm{p},L)$) or the permeability $\bm{p}$ as outputs. 

\subsection{\label{sec:vae}VAE and cVAE}

In the former section (see Sec.~\ref{sec:cnn}), the raw configuration $\bm{X}$ at different permeability $\bm{p}$ is generated by Monte Carlo simulation. However, what if configurations are not in the raw configurations, can we still identify the physical features as well? Here we consider to use the VAE and the cVAE to generate new configuration $\hat{\bm{X}}_{\text{VAE}}$ and $\hat{\bm{X}}_{\text{cVAE}}$, respectively. 

\begin{figure*}[htbp]
 \noindent\includegraphics[width=0.75\textwidth]{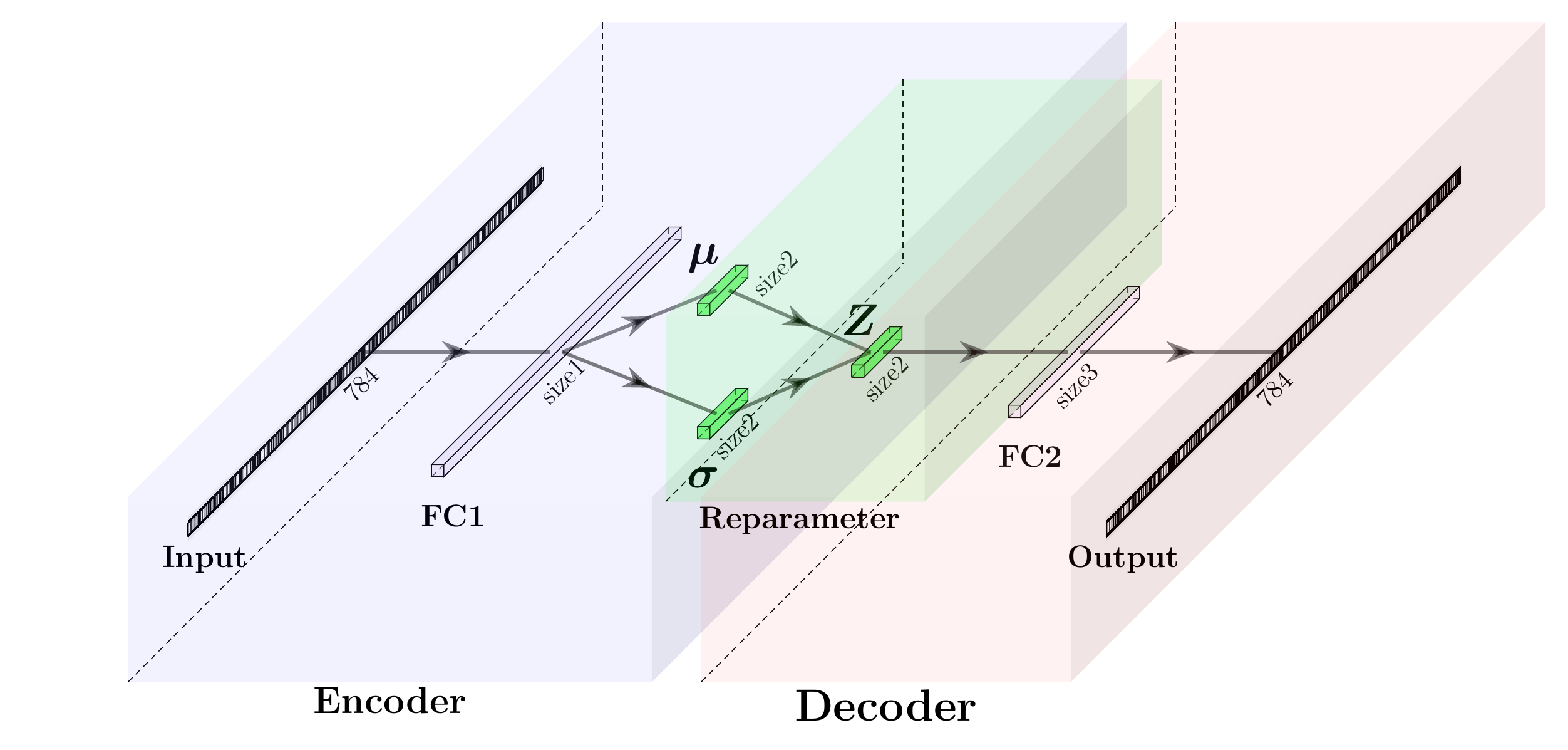}
 \caption{\label{fig:VAE} The structure of the VAE with an encoder and a decoder. The left large light purple cuboid refers to the encoder with two fully connected layers, i.e., ``FC1'',  and ``$\bm{\mu}$'' or ``$\bm{\sigma}$'', with the size of ``size1'', and ``size2''. And the right large light red cuboid is the decoder with two fully connected layers, including ``FC2'', and ``Output'' with the size of ``size3'', and 784. The outputs of the encoder are the mean value ``$\bm{\mu}$'' and the standard deviation ``$\bm{\sigma}$''. The input ``$\bm{Z}$'' of the decoder is sampled from the normal distribution with ``$\bm{\mu}$'' and ``$\bm{\sigma}$''. The green cuboid consists of ``$\bm{\mu}$'', ``$\bm{\sigma}$'', and ``$\bm{Z}$''. The rectangles with 784 black and white lattices represent percolating configuration $\bm{X}$ on the left and its reconstruction $\hat{\bm{X}}$ on the right, respectively.}
\end{figure*}

VAE (see Fig.~\ref{fig:VAE}), a generative network, bases on the variational Bayes inference proposed by \cite{Kingma2014Auto}. Contracted with traditional AE (see Fig. S. 1), the VAE describes latent variables with probability. From that point, the VAE shows great values in  data generation. Just like AE, the VAE is composed of an encoder and a decoder. The VAE uses two different CNNs as two probability density distributions. The encoder in the VAE, called the inference network $p_{\text{encoder}}(\bm{Z}|\bm{X})$, can generate the latent variables $\bm{Z}$. And the decoder in the VAE, called the generating network $p_{\text{decoder}}(\hat{\bm{X}}|\bm{Z})$, reconstructs the raw configuration $\bm{X}$. Unlike AE, the encoder and the decoder in VAE are constrained by the two probability density distributions.

\begin{figure*}[htbp]
 \noindent\includegraphics[width=0.9\textwidth]{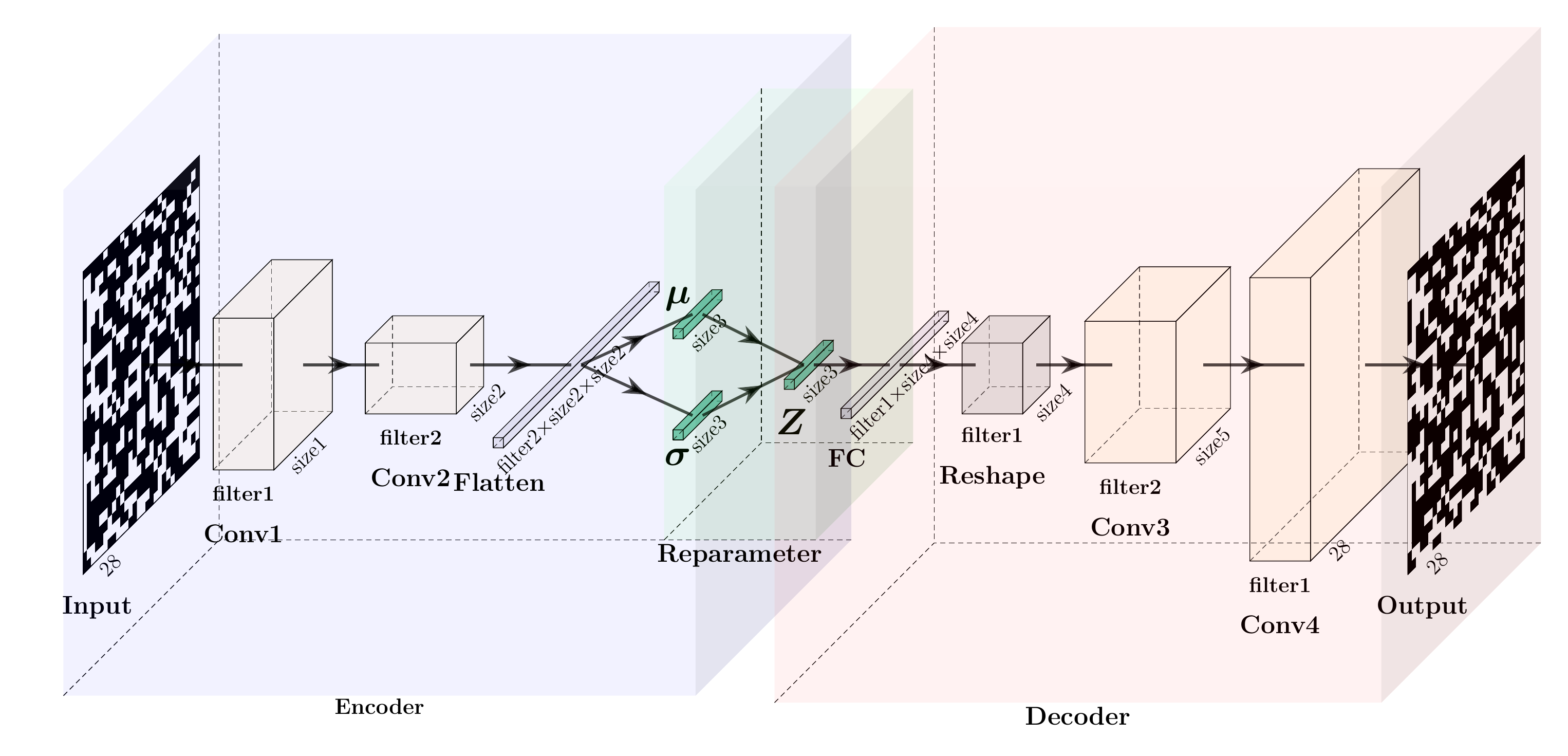}
 \caption{\label{fig:cVAE} The structure of the cVAE with an encoder and a decoder. The left large light purple cuboid refers to the encoder with three layers, including the input layer ``Input'' with the size of ``28$\times$28'', two convolution layers (``Conv1'' and ``Conv2'') with the size of ``filter1$\times$size1$\times$size1'' and ``filter2$\times$size2$\times$size2'', a flatten layer ``Flatten'' with the size of ``filter2$\times$size2$\times$size2'', and a fully connected layer (``$\bm{\mu}$'' or ``$\bm{\sigma}$'') with the size of ``size3''. And the right large light red cuboid is the decoder with four layers, including the layer ``$\bm{Z}$'' with the size of ``size3'', a fully connected layer ``FC'' with the size of ``filter1$\times$size4$\times$size4'', a reshape layer ``Reshape'' with the size of ``filter1$\times$size4$\times$size4'', two transposed convolution layers (``Conv3'' and ``Conv4'') with the size of ``filter2$\times$size5$\times$size5'' and ``filter1$\times$28$\times$28'', and a output layer ``Output'' with the size of ``28$\times$28''. The outputs of the encoder are the mean value ``$\bm{\mu}$'' and the standard deviation ``$\bm{\sigma}$''. The input of the decoder ``$\bm{Z}$'' is sampled from the normal distribution with ``$\bm{\mu}$'' and ``$\bm{\sigma}$''. The green cuboid is consist of ``$\bm{\mu}$'', ``$\bm{\sigma}$'', and ``$\bm{Z}$''. The squares with 784 black and white lattices represent percolating configurations $\{\bm{X}_1, \bm{X}_2, \ldots, \bm{X}_M\}$ on the left and their reconstructions $\{\hat{\bm{X}}_1, \hat{\bm{X}}_2, \ldots, \hat{\bm{X}}_M\}$ on the right, respectively. }
\end{figure*}

To better reconstruct the raw two-dimensional configuration $\bm{X}$, the fully connected layer in the VAE is replaced by the convolution layer. Now we have got the cVAE. The architecture of the cVAE is shown in Fig.~\ref{fig:VAE}. Usually, the performance of the cVAE is better than the VAE due to the configuration $\bm{X}$ with spatial attribute. In our work, the VAE and the cVAE are both used for generating THE new configuration $\hat{\bm{X}}$.

\subsection{\label{sec:pca}PCA}

The Sec.~\ref{sec:cnn} and Sec.~\ref{sec:vae} focus on supervised learning, which hypothesize that the labels exist for the raw configuration $\bm{X}$ on the percolation model. However, though we can detect the permeability values $\{0.41,0.42,\ldots,0.80\}$ and the two order parameters ($\varPi(\bm{p},L)$ and $P(\bm{p},L)$) by supervised learning, label dearth often occurs. Thus, it is imperative to identify the labels, such as $\varPi(\bm{p},L)$, $P(\bm{p},L)$, and $\bm{p}$. Some recent studies have shown that the first principal component obtained by PCA can be regarded as typical physical quantities \cite{Wang2016Discovering,wetzel2017unsupervised,hu2017discovering,yu2020unsupervised}. Base on these studies, we explore the meaning of the first principal component on the percolation model.

As it is well-known, PCA can reduce the dimension of the matrix $\bm{X}$. First, we compute the mean value $\bm{X}_\text{mean}=1/M\sum_{i=1}^{M}a_{ij} (i=1,2,\ldots,M;j=1,2,\ldots,N)$ for each column $\bm{C}_j (j=1,2,\ldots,N)$ in the matrix $\bm{X}$. Then we get the centered matrix $\bm{X}_\text{centered}$ that is expressed as $\bm{X}_\text{centered}=\bm{X}-\bm{X}_\text{mean}$. After obtaining $\bm{X}_\text{centered}$, by an orthogonal linear transformation expressed as $\bm{X}_\text{centered}^{T}\bm{X}_\text{centered}\bm{W}=\bm{\lambda}\bm{W}$, we extract the eigenvectors $\bm{W}$ and the eigenvalues $\bm{\lambda}$. The eigenvectors $\bm{W}$ are composed with $\bm{w}_1, \bm{w}_2,\ldots,\bm{w}_N$. The eigenvalues are sorted in the descending order, i.e., $\bm{\lambda}_1 \ge \bm{\lambda}_2 \ge \ldots \ge \bm{\lambda}_N \ge 0$. The normalized eigenvalues $\tilde{\bm{\lambda}}_{j} (j=1,2,\ldots,N)$ are expressed as $\bm{\lambda}_j/\sum_{j=1}^{N}{\bm{\lambda}_j}$. The row $\bm{R}_i$ in matrix $\bm{X}$ can be transformed into $\bm{X}^{'}_i=\bm{R}_{i}\bm{W}$. Eq.~\ref{eq:x'} represents the statistic average every 40 intervals for each permeability $p$. This process is quite similar to the process of calculating the two order parameters, i.e., $\varPi(\bm{p},L)$ and $P(\bm{p},L)$. Table~\ref{tab:algorithm} shows the procedure of PCA algorithm.
\begin{equation}
   \langle\bm{X}^{'}\rangle=\frac{1}{1000}\sum_{i=1}^{1000}|\bm{X}^{'}_{i\times 40}|
   \label{eq:x'}
\end{equation}
\begin{table}[b]
\caption{The procedure of PCA Algorithm}
\begin{ruledtabular}
\begin{tabular}{m{8.49cm}}
  \label{tab:algorithm}
  {Require: the raw configuration $\bm{X}$} \\    
  \hline    \\
  1. Compute the mean value $\bm{X}_\text{mean}$ for the column $\bm{C}_j$ in the matrix $\bm{X}$;  \\
  \vspace{1mm}
  2. Get the centered matrix $\bm{X}_\text{centered}$;\\
  \vspace{1mm}
  3. Compute the eigenvectors $\bm{W}$ and the eigenvalues $\bm{\lambda}$ by  an orthogonal linear transformation; \\
  \vspace{1mm}
  4. Transform $\bm{R}_i$ into $\bm{X}^{'}_i$;  \\
  \vspace{1mm}
  5. Get the statistic average $\langle\bm{X}^{'}\rangle$ with every 40 intervals for each permeability $p$. \\
  \vspace{1mm}
\end{tabular}
\end{ruledtabular}
\end{table}

\section{\label{sec:result_and_discussion} Results and Discussion}

\subsection{\label{sec:train_cnn} Simulate the two order parameters by two CNNs}
In this section, we consider to use the approches in Sec.~\ref{sec:machine_learning} to capture the physic features. First we make use of TensorFlow 2.2 library to perform the CNNs with four layers. To predict the two order parameters ($\varPi(\bm{p},L)$ and $P(\bm{p},L)$), two kinds of CNNs (CNNs-\uppercase\expandafter{\romannumeral1} and CNNs-\uppercase\expandafter{\romannumeral2}) are constructed. The first two layers of the former two kinds of CNNs are composed of two convolution layers (``Con1'' and ``Con2''), both of which possessed ``filter1''=32 and ``filter2''=64 filters with the size of $3\times 3$, and a stride of 1. Each convolution layer is followed by a max-pooling layer with the size of $2\times 2$. The final convolution layer ``Con2''  is strongly interlinked to a fully connected layer ``FC'' with 128 variables. The output layer ``Output'', following by ``FC'', is a fully connected layer. For the two convolution layers and the fully connected layer ``FC'', a rectified linear unit (ReLU) $\bm{a}=\text{max}(0,\bm{x})$ \cite{agarap2018deep} is chosen as activation function due to its reliability and validity. However, the output layer has no activation function.  

After determining the framework of CNNs-\uppercase\expandafter{\romannumeral1} and CNNs-\uppercase\expandafter{\romannumeral2}, here we mention how to train CNNs-\uppercase\expandafter{\romannumeral1} and CNNs-\uppercase\expandafter{\romannumeral2} for deducing $\varPi(\bm{p},L)$ and $P(\bm{p},L)$. First, we carry out an Adam algorithm \cite{kingma2014adam} as an optimizer to update parameters, i.e., weights and biases. Then, a mini batch size of 256 and a learning rate of $10^{-4}$ are selected for its timesaving. Following this treatment, CNNs-\uppercase\expandafter{\romannumeral1} and CNNs-\uppercase\expandafter{\romannumeral2} are trained on 1000 epochs for 40,000 uncorrelated and shuffled configurations, respectively. Before training, we split $\{\bm{X}_1,\bm{X}_2,\ldots,\bm{X}_M\}$, $\varPi(\bm{p},L)$ and $P(\bm{p},L)$ into 32,000 training set and 8,000 testing set. While training, we monitor three indicators, including the loss function (i.e., mean squared error (MSE, see Eq.~\ref{mse})), mean average error (MAE, see Eq.~\ref{mae}), and root mean squared error (RMSE, see Eq.~\ref{rmse}), for training and testing set \cite{cheng2020comparison}. In Eq.~\ref{mse}-\ref{rmse}, ${y_i}^{\text{raw}}$ and ${y_i}^{\text{pred}}$ refer to $\varPi(\bm{p},L)$/$P(\bm{p},L)$ and its predictions. 
If the loss function in testing set reaches the minimum, then the optimal CNNs-\uppercase\expandafter{\romannumeral1} and CNNs-\uppercase\expandafter{\romannumeral2} will be obtained. As can be seen from Fig. S. 2, these indicators gradually decrease. In Table. S. 3, the errors of the optimal CNNs-\uppercase\expandafter{\romannumeral1} and CNNs-\uppercase\expandafter{\romannumeral2} are very small. What stands out in Fig. S. 2 is that CNNs-\uppercase\expandafter{\romannumeral1} and CNNs-\uppercase\expandafter{\romannumeral2} have high stability, consistency, and faster convergence rate. 
\begin{equation}
   \mathrm{MSE}=\frac{1}{M}\sum_{i=1}^{M}({y_i}^{\text{pred}}-{y_i}^{\text{raw}})^2\
   \label{mse}
\end{equation}
\begin{equation}
   \mathrm{MAE}=\frac{1}{M}\sum_{i=1}^{M}|{y_i}^{\text{pred}}-{y_i}^{\text{raw}}|
   \label{mae}
\end{equation}
\begin{equation}
   \mathrm{RMSE}=\sqrt{\frac{1}{M}\sum_{i=1}^{M}({y_i}^{\text{pred}}-{y_i}^{\text{raw}})^2}
   \label{rmse}
\end{equation}

\begin{figure*}[htbp]
 \includegraphics[width=0.99\textwidth]{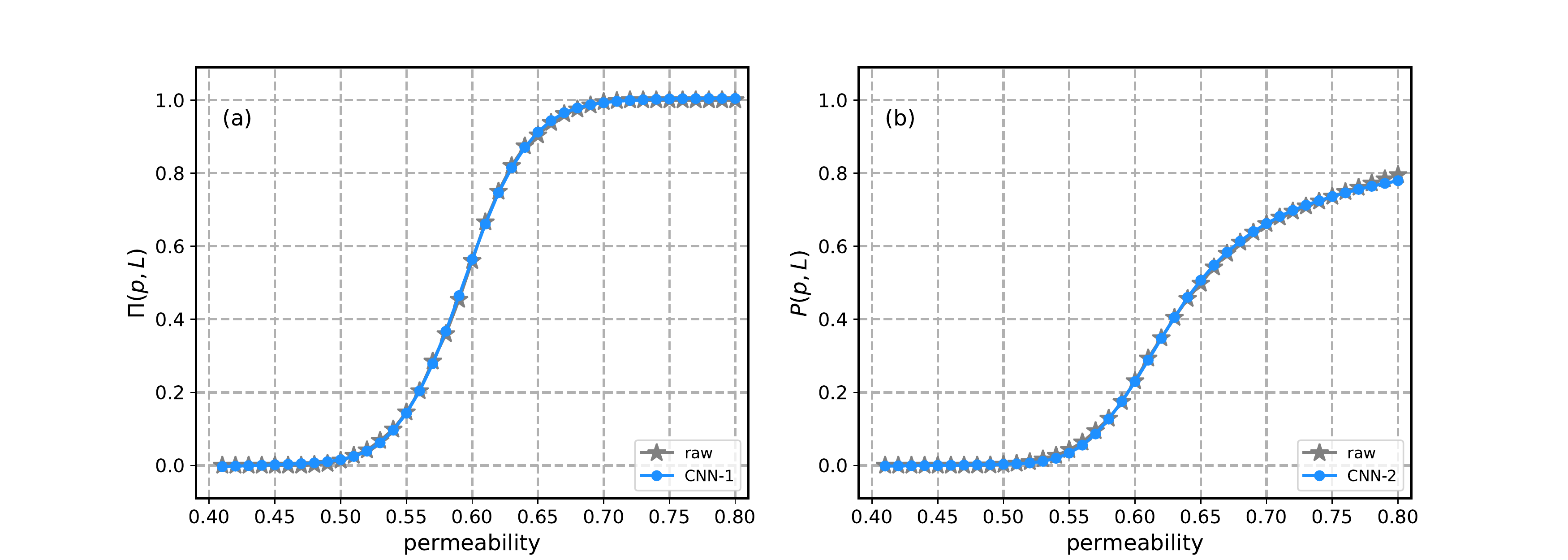}
 \caption{\label{fig:cnn-percolation-pi-p}(a) The relationship between the permeability values $\{0.41,0.42,\ldots,0.80\}$ and the raw $\varPi(\bm{p},L)$ or the statistic average from the outputs of CNNs-\uppercase\expandafter{\romannumeral1}. (b) The relationship between the permeability values $\{0.41,0.42,\ldots,0.80\}$ and the raw $P(\bm{p},L)$ or the statistic average from the outputs of CNNs-\uppercase\expandafter{\romannumeral2}.}
 
\end{figure*}

Before assess CNNs-\uppercase\expandafter{\romannumeral1} and CNNs-\uppercase\expandafter{\romannumeral2}, we have to explain what is meant by statistic average. Statistic average can be defined as the averages of CNNs-\uppercase\expandafter{\romannumeral1}'s or CNNs-\uppercase\expandafter{\romannumeral2}'s outputs for each permeability $p$. As shown in Fig.~\ref{fig:cnn-percolation-pi-p}, there is a clear trend of phase transition between the permeability values $\{0.41,0.42,\ldots,0.80\}$ and $\varPi(\bm{p},L)$/$P(\bm{p},L)$. The two grey lines in Fig.~\ref{fig:cnn-percolation-pi-p}, which are the same as the two blue lines in Fig.~\ref{fig:percolation-pi-p}, represent the relationship between the permeability values $\{0.41,0.42,\ldots,0.80\}$ and the raw $\varPi(\bm{p},L)$ or $P(\bm{p},L)$. Likewise, the two blue lines in Fig.~\ref{fig:cnn-percolation-pi-p}, represent the relationship between the permeability values $\{0.41,0.42,\ldots,0.80\}$ and the statistic average from the outputs of CNNs-\uppercase\expandafter{\romannumeral1} and CNNs-\uppercase\expandafter{\romannumeral2}. The overlapping of the two kinds of lines shows that CNNs-\uppercase\expandafter{\romannumeral1} and CNNs-\uppercase\expandafter{\romannumeral2} can be used to deduce the two order parameters and the process of phase transition.

\begin{figure}[htbp]
 \includegraphics[width=0.45\textwidth]{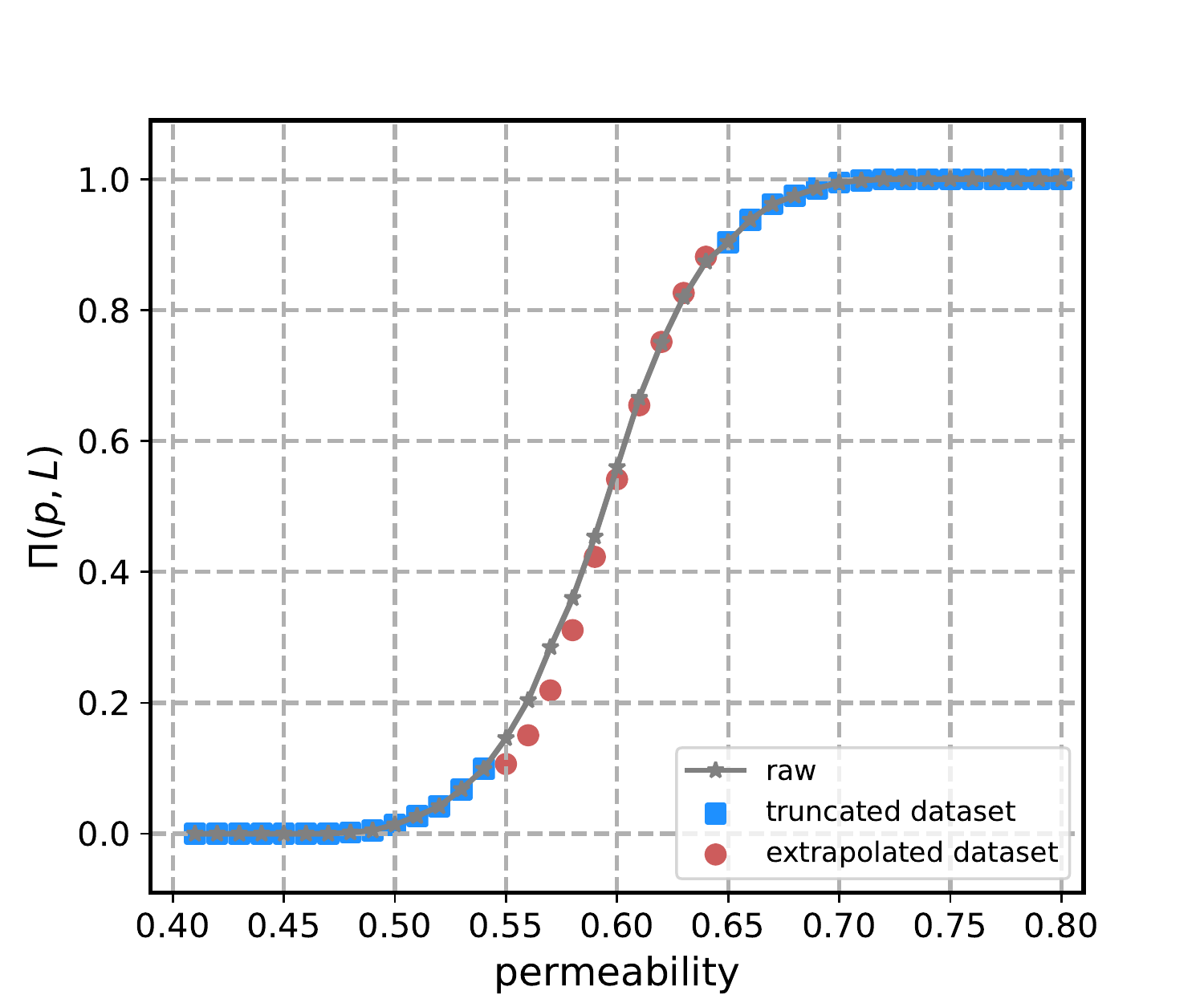}
 \caption{\label{fig:cnn-percolation-pi-inference} Identification of phase transition with truncated dataset by CNNs-\uppercase\expandafter{\romannumeral1}. The permeability values $\{0.41,0.42,\ldots,0.80\}$ and $\varPi(\bm{p},L)$ connected with red points are artificially removed from the dataset. And CNNs-\uppercase\expandafter{\romannumeral1} makes the judgment only by learning the data associated with blue points. The grey curve show that $\varPi(\bm{p},L)$ shifts with the permeability values $\{0.41,0.42,\ldots,0.80\}$.}
\end{figure}

To overcome the difficulty associated with the percolation model being near the critical transition point, we truncate the dataset. Specifically, we remove the data near the critical transition point, and only retain the data far away from the critical transition point. Here we take the simulation of $\varPi(\bm{p},L)$ as an example. The retained data with the raw $\varPi(\bm{p},L)$ rangs from 0 to 0.1, and 0.9 to 1. As shown in Fig. S. 3 and the middle red points in Fig.~\ref{fig:cnn-percolation-pi-inference}, we find that CNNs-\uppercase\expandafter{\romannumeral1} can extrapolate $\varPi(\bm{p},L)$ to missing data by learning the retained data.

\begin{figure*}[htbp]
 \includegraphics[width=0.99\textwidth]{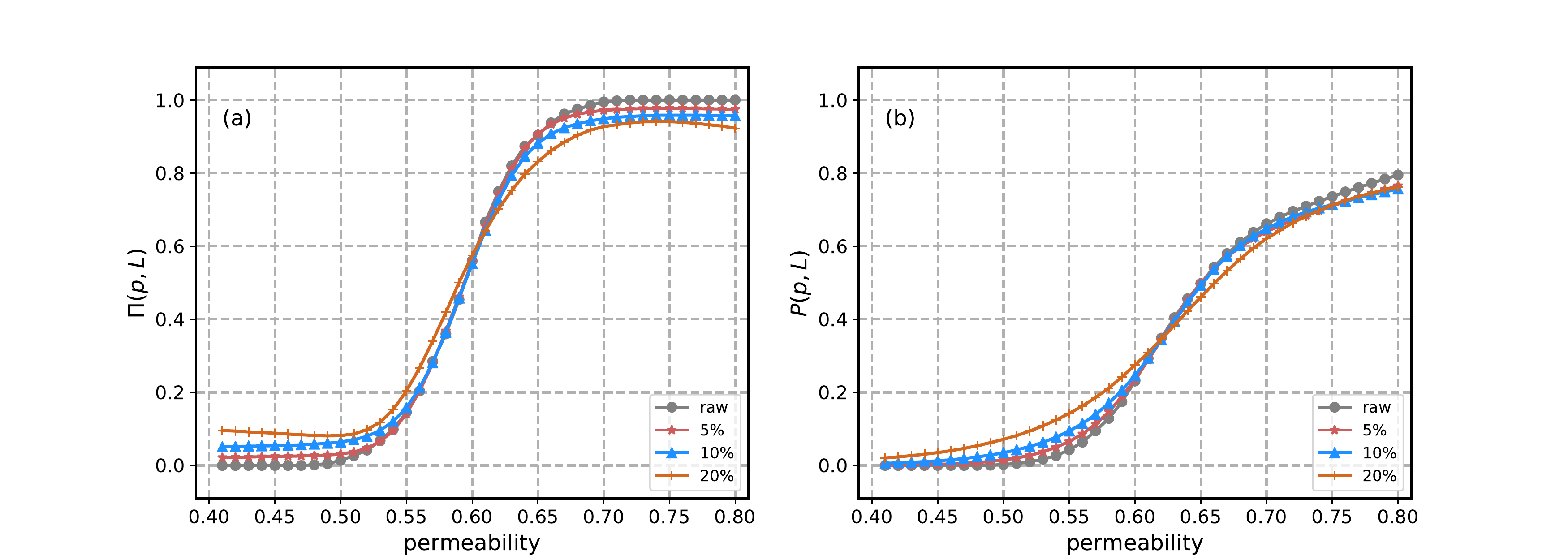}
 \caption{\label{fig:cnn-percolation-pi-p-noise}Robustness of the two order parameters ($\varPi(\bm{p},L)$ and $P(\bm{p},L)$) with noises for CNNs-\uppercase\expandafter{\romannumeral1} and CNNs-\uppercase\expandafter{\romannumeral2}. (a) The relationship between the permeability values $\{0.41,0.42,\ldots,0.80\}$ and the raw $\varPi(\bm{p},L)$ or the statistic average from the outputs of CNNs-\uppercase\expandafter{\romannumeral1} under different noisy inputs. (b) The relationship between the permeability values $\{0.41,0.42,\ldots,0.80\}$ and the raw $P(\bm{p},L)$ or the statistic average from the outputs of CNNs-\uppercase\expandafter{\romannumeral2} under different noisy inputs.}
\end{figure*}

This section demonstrate that CNNs-\uppercase\expandafter{\romannumeral1} and CNNs-\uppercase\expandafter{\romannumeral2} can be two effective tools for detecting $\varPi(\bm{p},L)$ and $P(\bm{p},L)$, respectively. Additional test should be made to verify that whether or not CNNs-\uppercase\expandafter{\romannumeral1} and CNNs-\uppercase\expandafter{\romannumeral2} are robust against noise. To address this issue, we deliberately invert a proportion, i.e., 5\%, 10\%, and 20\%, of the labels for the raw $\varPi(\bm{p},L)$ and $P(\bm{p},L)$ and verify that whether or not the “artificial” noises can affect the predicted $\varPi(\bm{p},L)$ and $P(\bm{p},L)$. Fig.~\ref{fig:cnn-percolation-pi-p-noise} and Fig. S. 4 demonstrate that CNNs-\uppercase\expandafter{\romannumeral1} and CNNs-\uppercase\expandafter{\romannumeral2} are robust against noise. As the labeling error rates increase, the same trend is evident in the outputs of CNNs-\uppercase\expandafter{\romannumeral1} and CNNs-\uppercase\expandafter{\romannumeral2} within a relatively small difference (see Fig.~\ref{fig:cnn-percolation-pi-p-noise}). Therefore, we draw the conclusion that noises have little effect on detecting $\varPi(\bm{p},L)$ and $P(\bm{p},L)$.
 
\subsection{\label{sec:train_cnn_percolation}Simulate the permeability by one CNNs}

Just like we use CNNs-\uppercase\expandafter{\romannumeral1} and CNNs-\uppercase\expandafter{\romannumeral2} in Sec.~\ref{sec:train_cnn}, here we use the same structure for CNNs-\uppercase\expandafter{\romannumeral3} and strategies for training the permeability $\bm{p}$. The only distinction between CNNs-\uppercase\expandafter{\romannumeral3} and CNNs-\uppercase\expandafter{\romannumeral1}/CNNs-\uppercase\expandafter{\romannumeral2} is that the outputs for CNNs-\uppercase\expandafter{\romannumeral3} are the permeability $\bm{p}$ instead of the two order parameters. Fig. S. 5 shows the performance of CNNs-\uppercase\expandafter{\romannumeral3}. With successive increases in epochs, the MSE, MAE, and RMSE continue to decrease until no longer dropping. 
 
\begin{figure}[htbp]
 \noindent\includegraphics[width=0.48\textwidth]{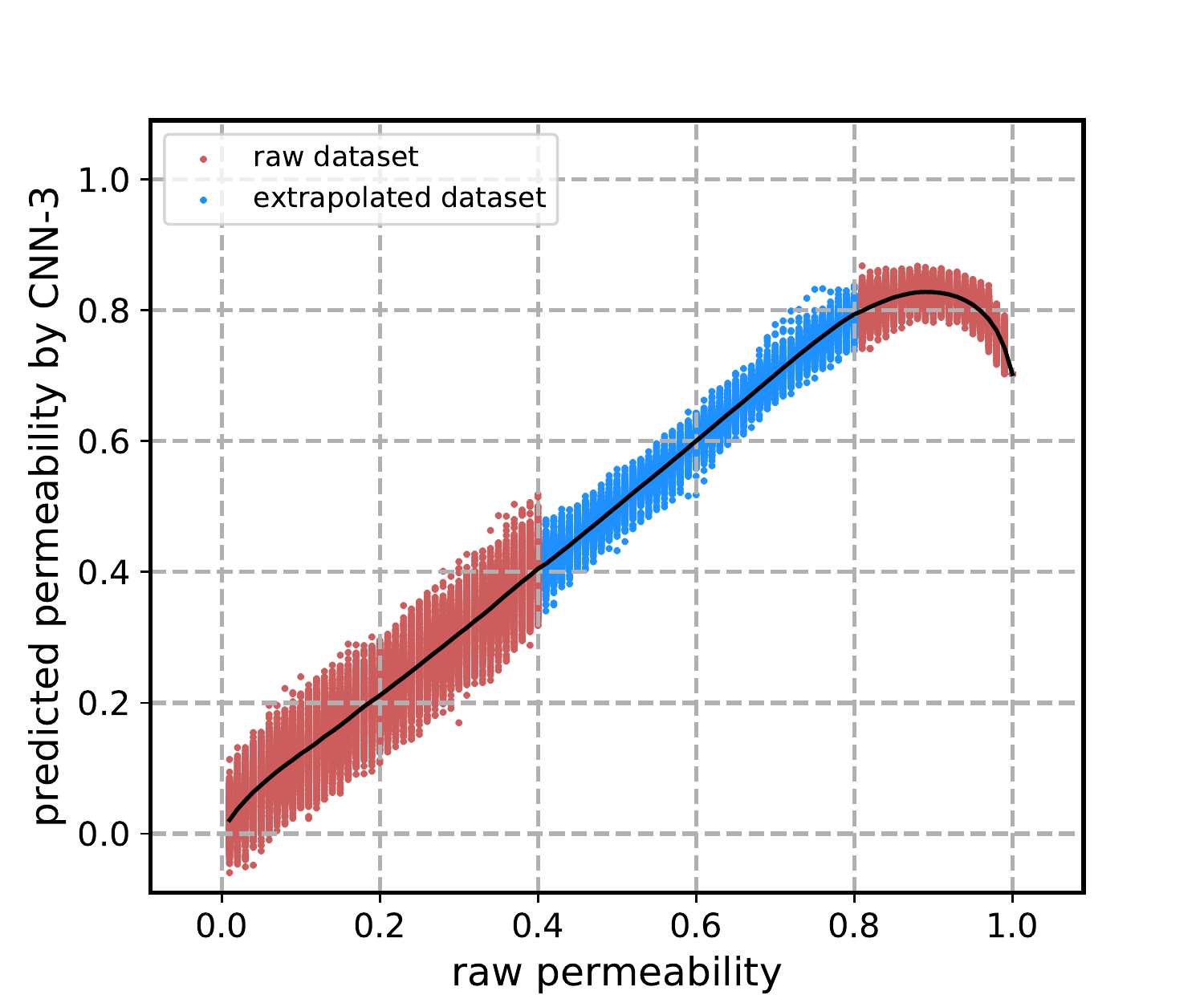}
 \caption{\label{fig:cnn-percolation-inference}The correlation between the raw permeability $\bm{p}$ and its predictions. The blue circles refer to the relationship between the raw permeability $\bm{p}$ in dataset and its predictions $\hat{\bm{p}}$ by CNNs-\uppercase\expandafter{\romannumeral3}. The red circles refer to the relationship between the raw permeability $\bm{p}_{\text{ex}}$ out of dataset and its prediction $\hat{\bm{p}}_{\text{extrapolated}}$ by CNNs-\uppercase\expandafter{\romannumeral3}. The raw permeability $\bm{p}_{\text{ex}}$ out of dataset not only range from 0.01 to 0.4, and from 0.81 to 1.0, with an interval of 0.01. The black line refers to the correlation between the raw permeabilities $\{\bm{p},\bm{p}_{\text{extrapolated}}\}$ ranges from 0.01 to 1.0 with an interval of 0.01 and their average predictions.}
\end{figure}

Another measure of CNNs-\uppercase\expandafter{\romannumeral3}'s performance is concerned with the difference between the raw permeability $\bm{p}$ and its prediction $\hat{\bm{p}}$. The blue circles in Fig.~\ref{fig:cnn-percolation-inference} show that there is a strong positive correlation between the raw permeability $\bm{p}$ and its prediction $\hat{\bm{p}}$. Further statistical tests reveal that most of the gap between the raw permeability $\bm{p}$ and its prediction $\hat{\bm{p}}$ is less than 0.1. The result proves that CNNs-\uppercase\expandafter{\romannumeral3} has the advantage of convenient use and high precision.

On the other hand, extrapolation ability can also reflect the performance of CNNs-\uppercase\expandafter{\romannumeral3}. To do this, we use Monte Carlo simulation to generate new dataset. The new permeability $\bm{p}$ not only ranges from 0.01 to 0.40, but also from 0.81 to 1.00, with an interval of 0.01. Just like Sec.~\ref{sec:percolation}, we perform 1050 Monte Carlo steps and keep the last 1000 steps. As a consequence, 60,000 configurations are generated. The red circles in Fig.~\ref{fig:cnn-percolation-inference} exhibit the extrapolation ability of CNNs-\uppercase\expandafter{\romannumeral3}. Our results show that no significant correlation between the extrapolated permeability $\bm{p}_{\text{ex}}$ ranging from 0.81 to 1.00 and their prediction $\hat{\bm{p}}_{\text{extrapolated}}$. However, in Fig.~\ref{fig:cnn-percolation-inference}, the results indicate that CNNs-\uppercase\expandafter{\romannumeral3} has good extrapolation ability for the extrapolated permeability $\bm{p}_{\text{ex}}$ from 0.01 to 0.40.

\subsection{\label{sec:train_ae}Generate new configurations by one VAE and one cVAE}

Though CNNs-\uppercase\expandafter{\romannumeral1}, CNNs-\uppercase\expandafter{\romannumeral2}, and CNNs-\uppercase\expandafter{\romannumeral3} are valid when detecting the two order parameters ($\varPi(\bm{p},L)$ and $P(\bm{p},L)$) and the permeability $\bm{p}$, the validity of these three CNNs is unkown for percolating configurations outside of the dataset. As is shown in Fig.~\ref{fig:VAE} and Fig.~\ref{fig:cVAE}, we use the same network structures for the VAE and the cVAE to generate new configurations ($\hat{\bm{X}}_{\text{vae}}$ and $\hat{\bm{X}}_{\text{cvae}}$). Actually, $\hat{\bm{X}}_{\text{vae}}$ and $\hat{\bm{X}}_{\text{cvae}}$ can be regarded as adding some noise into the raw configuration $\bm{X}$.

Let us first consider VAE. Just like AE (see Fig. S. 1), the VAE is also composed of an encoder and a decoder. The encoder of the VAE owns two fully connected layers, both of which follows with a ReLU activation function. The first layer of the encoder possess ``size1''=512 neurons. Another 512 neurons, including 256 mean ``$\bm{\mu}$'' and 256 variance ``$\bm{\sigma}$'', are taken into account for the second layer of the encoder. By resampling from the Gaussian distribution with the mean ``$\bm{\mu}$'' and the variance ``$\bm{\sigma}$'', we obtain 256 latent variables ``$\bm{Z}$'' which are the inputs of the decoder. For the decoder of the VAE, two fully connected layers follow with the outputs of the encoder. For symmetry, the first layer of the decoder also contains ``size1''=512 neurons and follows with a ReLU activation function. And the output layer of the decoder contains 784 neurons which are used to reconstruct the raw configuration $\bm{X}$. Thus, the neurons in the output layer are the same as that in the input layer. 

Moving on now to consider cVAE. The encoder of the cVAE is composed of one input layer, two hidden convolution layers with ``filter1''=32 and ``filter2''=64 filters with the size of 3 and a stride of 2, and a ReLU activation function. The output layer in the encoder is a fully connected flatten layer with 800 neurons (400 mean ``$\bm{\mu}$'' and 400 variance ``$\bm{\sigma}$'') without activation function. By resampling from the Gaussian distribution with the mean ``$\bm{\mu}$'' and the variance ``$\bm{\sigma}$'', we obtain 400 latent variables ``$\bm{Z}$''. The reason why latent variables in the cVAE is more than the VAE is that the cVAE needs to consider more complex spatial characteristic. The decoder of the cVAE is composed of an input layer with 400 latent variables ``$\bm{Z}$''. A fully connected layer ``FC'' with 1,568 neurons is followed by ``$\bm{Z}$''. After reshaping the outputs of ``FC'' into three dimension, we feed the data into two transposed convolution layers (``Conv3'' and ``Conv4'') and one output layer ``Output''. The filters in these deconvolution layers are 64, 32 and 1 with the size of 3 and the stride of 2. After excluding the output layer ``Output'' without activation functions, there exist two ReLU activation functions followed by ``Conv3'' and ``Conv4'', respectively. 

We train the VAE and the cVAE over $10^3$ epochs using the Adam optimizer, a learning rate of $10^{-3}$, and a mini batch size of 256. To train the VAE and the cVAE, we use the sum of binary cross-entropy (see Eq.~\ref{eq:BinaryCrossEntropy}) and the Kullback-Leibler (KL) divergence (see Eq.~\ref{eq:kl}) as the loss function \cite{d2020learning}. In Eq.~\ref{eq:BinaryCrossEntropy}-\ref{eq:kl}, $\bm{x}_i^{\text{raw}}$ and $\bm{x}_i^{\text{pred}}$ represent each raw configuration with one/two dimension and its prediction. As shown in Fig. S. 6, the loss function, the binary cross-entropy, and the KL divergence vary with epochs for the VAE and the cVAE. Here we focus on the minimum value of loss function. From Fig. S. 6, the optimal cVAE performs better than the optimal VAE.
\begin{align}
\text{BinaryCrossEntropy}=-\sum_{i=1}^{M}((\bm{x}_i^{\text{pred}}\times \text{log}(\bm{x}_i^{\text{raw}})\notag\\
+(1-\bm{x}_i^{\text{pred}})\times \text{log}(1-\bm{x}_i^{\text{raw}}))
\label{eq:BinaryCrossEntropy}.
\end{align}
\begin{equation}
\text{KL}_{\text{divergence}}=-\sum_{i=1}^{M}\Biggl(\bm{x}_i^{\text{raw}}\times \text{log}\left(\frac{\bm{x}_i^{\text{raw}}}{\bm{x}_i^{\text{pred}}}\right)\Biggr)
\label{eq:kl}.
\end{equation}

\begin{figure}[htbp]
    \noindent\includegraphics[width=0.48\textwidth]{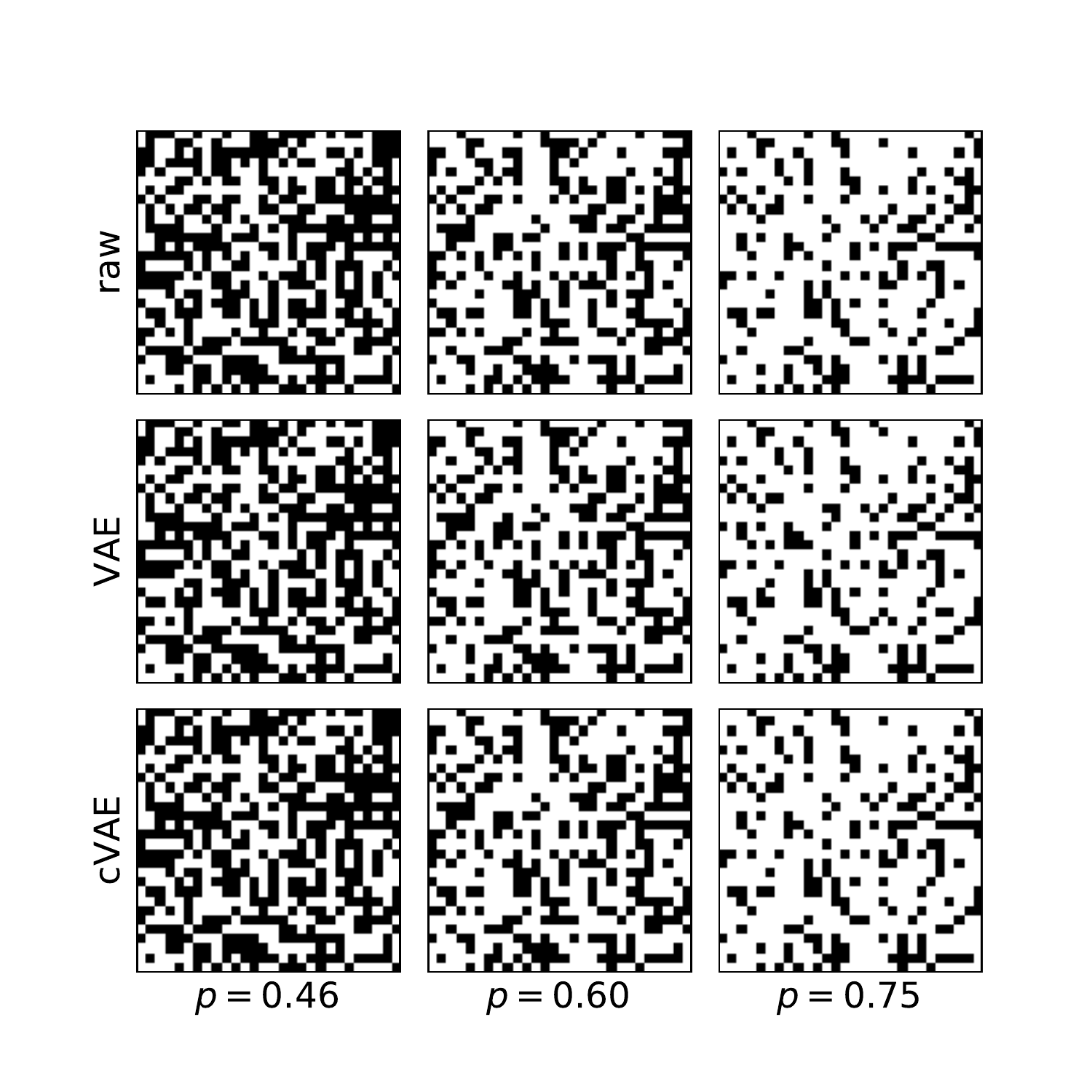}
    \caption{\label{fig:vae-cvae-image} Snapshots of percolating configurations for $p\in\{0.46,0.60,0.75\}$. The configurations in the top, middle, and bottom panels are sampled from the Monte Carlo simulation, the VAE, and the cVAE, respectively.}
\end{figure}

For a more visual comparison, we show the snapshots of the raw configurations $\{\bm{X}_1,\bm{X}_2,\ldots,\bm{X}_M\}$, and compare them to the VAE-generated and cVAE-generated configurations in Fig.~\ref{fig:vae-cvae-image}. As we can see from Fig.~\ref{fig:vae-cvae-image}, the configurations from the Monte Carlo simulation, the VAE and the cVAE are very close to each other. 

\begin{figure*}[htbp]
  \noindent\includegraphics[width=0.99\textwidth]{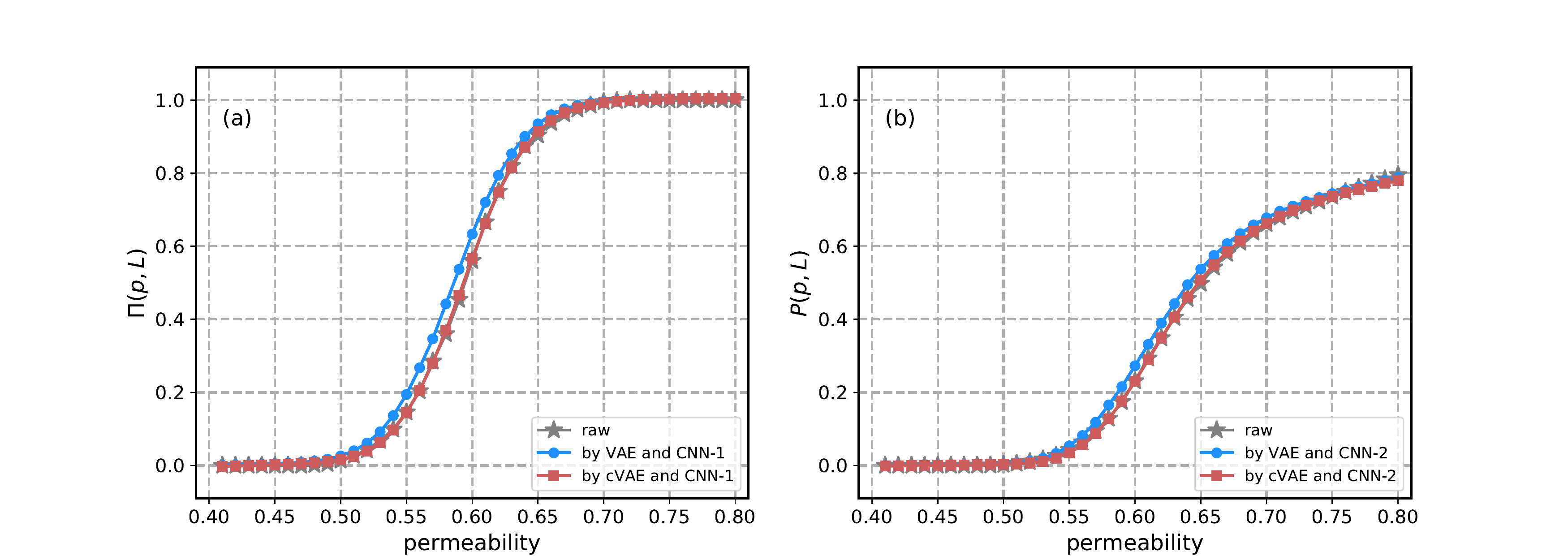}
  \caption{\label{fig:vae-cvae-percolation-pi-p}The relationship between the permeability values $\{0.41,0.42,\ldots,0.80\}$ and the $\varPi(\bm{p},L)$ or the statistic averages from the outputs of CNNs-\uppercase\expandafter{\romannumeral1} that originate from the outputs of the VAE and the cVAE. (b) The relationship between the permeability values $\{0.41,0.42,\ldots,0.80\}$ and $P(\bm{p},L)$ or the statistic averages from the outputs of CNNs-\uppercase\expandafter{\romannumeral2} that originate from the outputs of the VAE and the cVAE.}
\end{figure*}

\begin{figure*}[htbp]
 \noindent\includegraphics[width=0.90\textwidth]{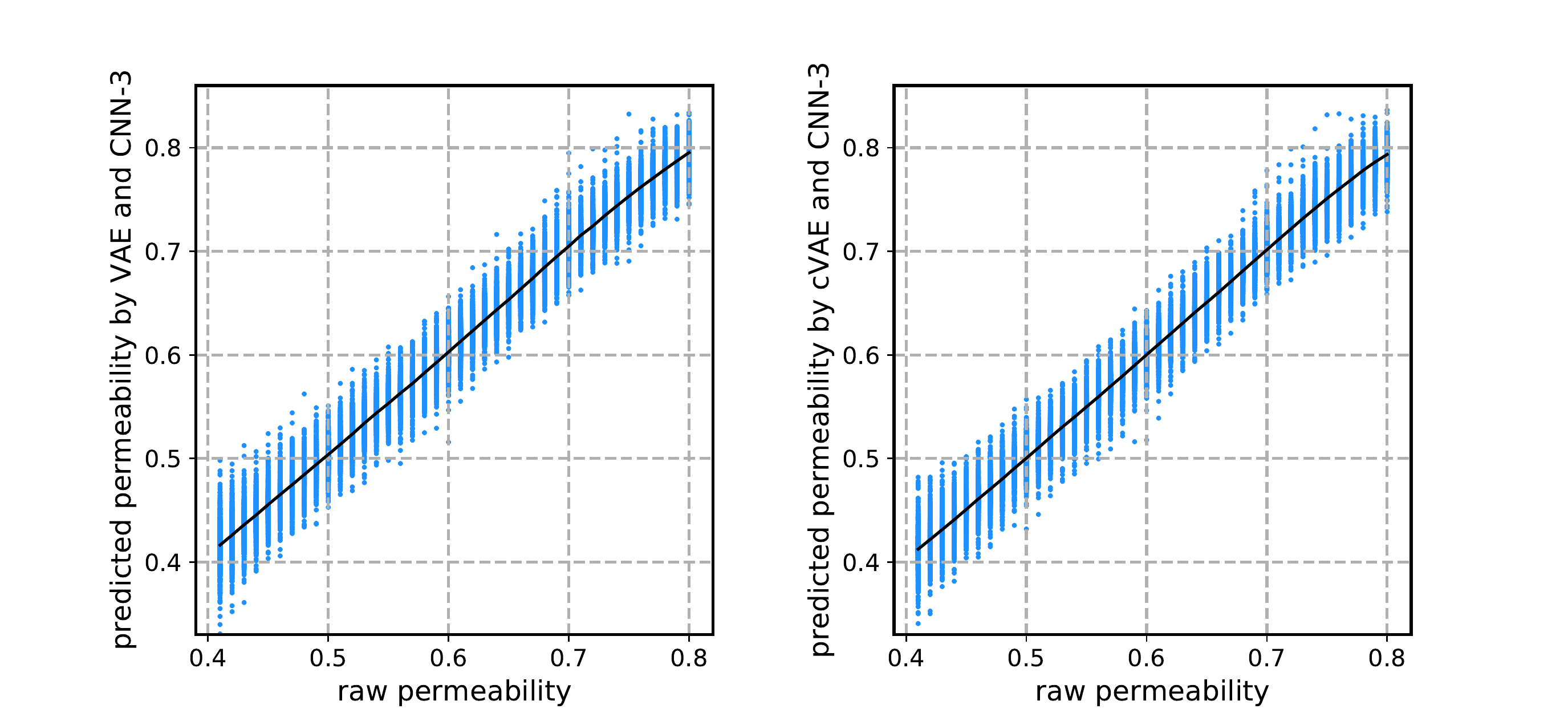}
 \caption{\label{fig:vae-cvae-percolation-inference}The correlation between the raw permeability $\bm{p}$ and its predictions through VAE/cVAE and CNNs-\uppercase\expandafter{\romannumeral3}. The blue circles refers to the relationship between the raw permeability $\bm{p}$ and its predictions by VAE/cVAE and CNNs-\uppercase\expandafter{\romannumeral3}. And the black line refers to the correlation between the raw permeability $\bm{p}$ and its average predictions.}
\end{figure*}

After reconstructing the configurations ($\hat{\bm{X}}_{\text{vae}}$ and $\hat{\bm{X}}_{\text{cvae}}$) through the VAE and the cVAE and pouring $\hat{\bm{X}}_{\text{vae}}$ and $\hat{\bm{X}}_{\text{cvae}}$ into CNNs-\uppercase\expandafter{\romannumeral1}, CNNs-\uppercase\expandafter{\romannumeral2}, and CNNs-\uppercase\expandafter{\romannumeral3}, we can detect $\varPi(\bm{p},L)$, $P(\bm{p},L)$, and the permeability $\bm{p}$. Fig.~\ref{fig:vae-cvae-percolation-pi-p} shows the relationships between the permeability values $\{0.41,0.42,\ldots,0.80\}$ and the statistic average from the outputs in CNNs-\uppercase\expandafter{\romannumeral1} and CNNs-\uppercase\expandafter{\romannumeral2}, respectively. From the red and purple lines in Fig.~\ref{fig:vae-cvae-percolation-pi-p}, by using $\hat{\bm{X}}_{\text{vae}}$ and $\hat{\bm{X}}_{\text{cvae}}$, we can obtain the two order parameters as well. From the Fig.~\ref{fig:vae-cvae-percolation-inference}, the raw permeability $\bm{p}$ is remarkably correlated linearly with its prediction $\hat{\bm{p}}$ through the VAE/cVAE and CNNs-\uppercase\expandafter{\romannumeral3}. Thus, subtle change in the raw configurations does not effect the catch of physical features. 

\subsection{\label{sec:train_pca}Identify the characteristic of the first principal component by one PCA}

This section will discuss how to capture the physic characteristics without labels. Various studies suggest to use PCA to identify order parameters \cite{Wang2016Discovering,wetzel2017unsupervised,hu2017discovering,yu2020unsupervised}. Therefore, we try to verify the feasibility of PCA in capturing order parameter on the percolation model.   

\begin{figure}[htbp]
    \noindent\includegraphics[width=0.48\textwidth]{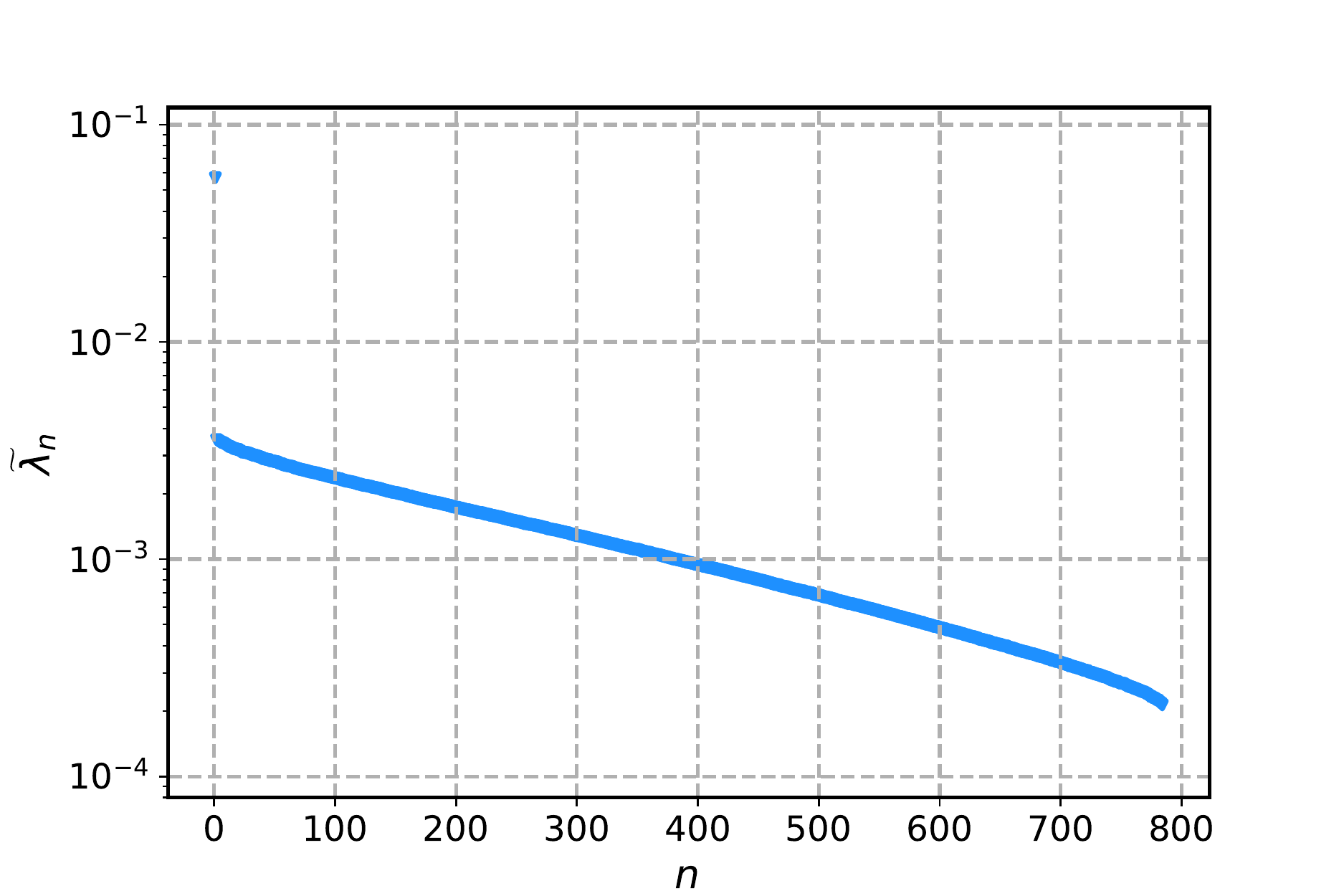}
    \caption{\label{fig:eigenvalues}The explained variance ratios obtained from the raw configuration $\bm{X}$ by the PCA, with the horizontal axis indicating corresponding component labels. The largest value of the explained variance ratios locating at the top-left corner means that there exists one dominant principle component.}
\end{figure}

First, we perform the PCA to the raw configuration $\bm{X}$. Fig.~\ref{fig:eigenvalues} exhibit the $N$ normalized eigenvalues $\tilde{\bm{\lambda}}_n=\bm{\lambda}_n/\sum_{n=1}^{N}\bm{\lambda}_n$. $\tilde{\bm{\lambda}}_n$ is also called as the explained variance ratios. The most noteworthy information in Fig.~\ref{fig:eigenvalues} is that there is one dominant principal component $\tilde{\bm{\lambda}}_1$, whcih is the largest one among $\tilde{\bm{\lambda}}_n$ and much larger than other explained variance ratios. Thus, $\tilde{\bm{\lambda}}_1$ plays a key role when dealing with dimension reduction. Based on $\tilde{\bm{\lambda}}_1$, the raw configuration $\bm{X}$ are mapped to another matrix $\bm{Y}=\bm{X}\tilde{\bm{\lambda}}_1$.

\begin{figure}[htbp]
    \noindent\includegraphics[width=0.48\textwidth]{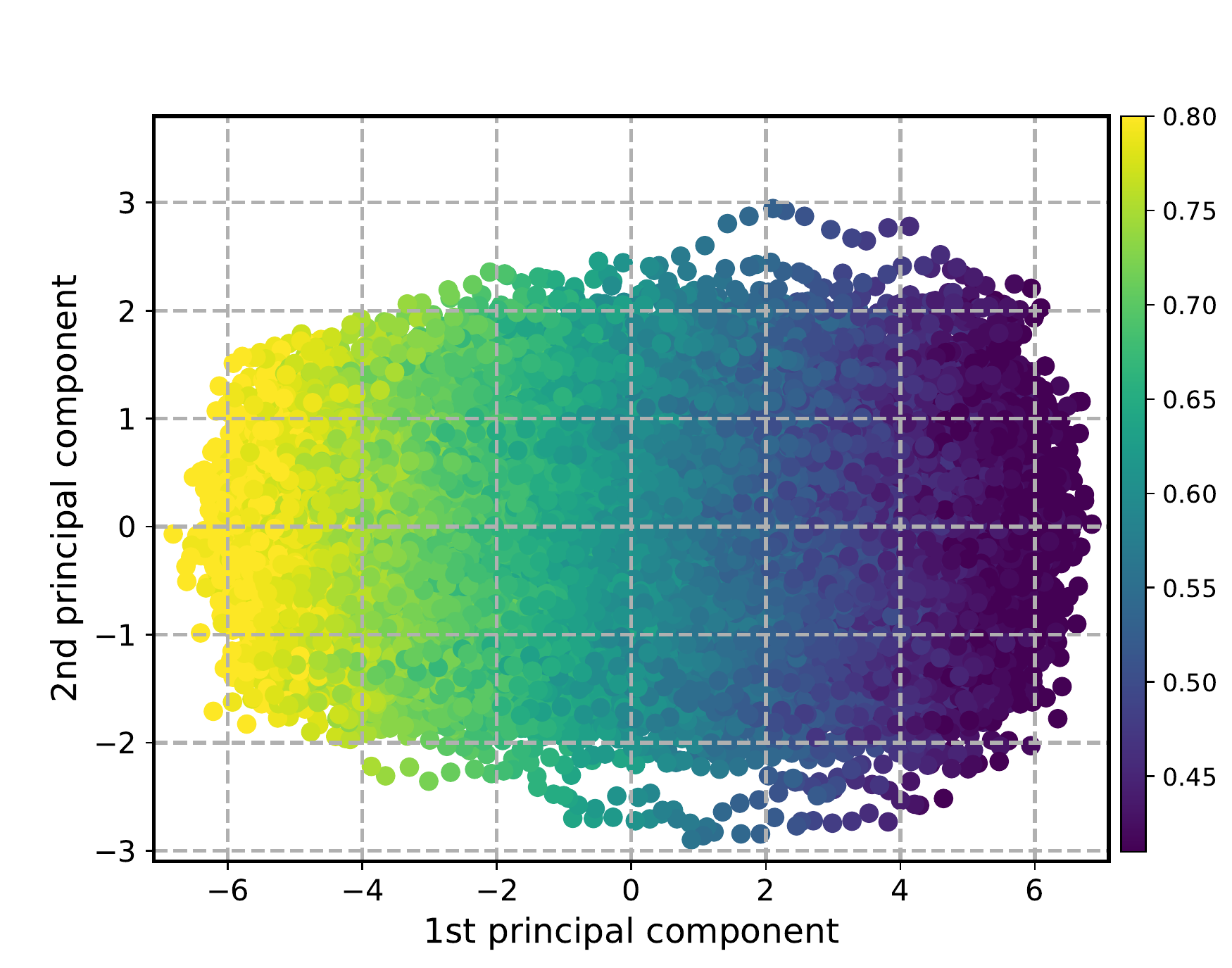}
    \caption{\label{fig:1st-2nd} Projection of the raw configuration $\bm{X}$ onto the plane of the first two dominant principal components, i.e., $\bm{X}\tilde{\bm{\lambda}}_1$ and $\bm{X}\tilde{\bm{\lambda}}_2$. The color bar on the right indicates the permeability values $\{0.41,0.42,\ldots,0.80\}$.}
\end{figure}

In Fig~\ref{fig:1st-2nd}, we construct the matrix $\bm{Y}^{'}=\{{\bm{X}\tilde{\bm{\lambda}}_1,\bm{X}\tilde{\bm{\lambda}}_2}\}$ by the first two eigenvalues and their eigenvectors. We use 40,000 blue scatter points to plot the the relationship between $\bm{X}\tilde{\bm{\lambda}}_1$ and $\bm{X}\tilde{\bm{\lambda}}_2$ on 40 permeability values ranging from 0.41 to 0.8. Just like in Fig.~\ref{fig:eigenvalues}, there is only one dominant representation on the percolation model due to the first principal component is much more important than the second principal component. 

\begin{figure*}[htbp]
 \noindent\includegraphics[width=0.99\textwidth]{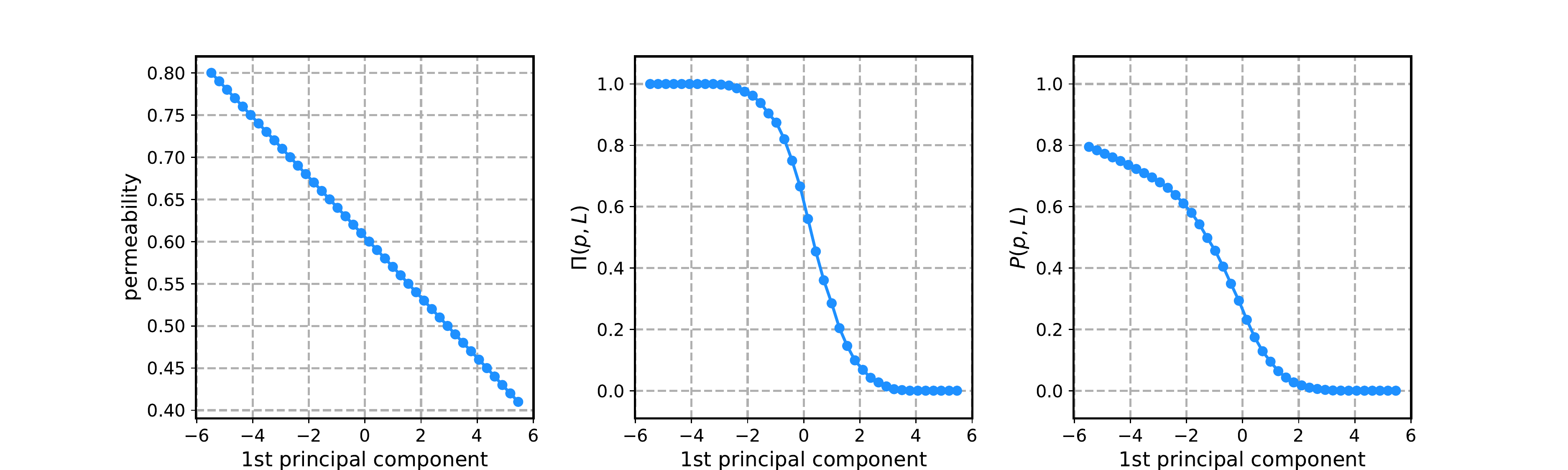}
 \caption{\label{fig:1st} Taking the normalized quantified first principal component $\bm{X}\tilde{\bm{\lambda}}_1$ as a function of the permeability $\bm{X}$ and the two order parameters, i.e., $\varPi(\bm{p},L)$ and $P(\bm{p},L)$.}
\end{figure*}

Having analysed the importance of the first principal component, we now move on to discuss the meaning of the first principal component. In Fig.~\ref{fig:1st}, we focus on the quantified first principal component as a function of the permeability $\bm{p}$ and the two order parameters, i.e., $\varPi(\bm{p},L)$ and $P(\bm{p},L)$. From Fig.~\ref{fig:1st}, we can see that there is a strong linear correlation between the quantified first principal component and the permeability $\bm{p}$. And the relationships between the quantified first principal component and two order parameters ($\varPi(\bm{p},L)$ and $P(\bm{p},L)$) are similar to the relationships between the permeability values $\{0.41,0.42,\ldots,0.80\}$ and the two order parameters. Our results are significant different from former study which demonstrates that the quantified first principal component can be taken as order parameter by data preprocessing \cite{yu2020unsupervised}. A possible explanation may be that \cite{yu2020unsupervised} evaluates the first principal component by removing the raw dataset with certain attributes on the percolation model. Therefore, we assume that the quantified first principal component obtained by PCA may not be taken as order parameter for various physical models. Another possible explanation is that, under certain conditions, the quantified first principal component can be regarded as order parameter.

\subsection{\label{sec:train_kmeans_cnn}Identify physic characteristics by one $k$-means and one CNNs}

From Sec.~\ref{sec:train_pca}, no significant corresponding is found between the normalized quantified first principal component and the two order parameters. Here we eagerly wonder how to identify order parameter. And another physical characteristics we desire to explore is the critical transition point. So we try to find a way to capture order parameter from the raw configurations $\{\bm{X}_1,\bm{X}_2,\ldots,\bm{X}_M\}$ and their permeability $\bm{p}$ on the percolation model. 

As it is well-known, for the two-dimensional percolating configurations, the critical transition point is equal to 0.593 in theoretical calculation. Though the critical transition point is already known, we wonder that whether or not the critical transition point can be found by machine. To do this, we first use a cluster analysis algorithm named $k$-means \cite{hartigan1979algorithm} to separate the raw configurations $\{\bm{X}_1,\bm{X}_2,\ldots,\bm{X}_M\}$ into two categories. The minimum and maximum value of their permeability $\bm{p}$ are 0.55 and 0.80 in the first cluster, and 0.41 and 0.65 in the second cluster. Note that there is an overlapping interval between 0.55 and 0.65 for the two categories. According to the overlapping interval, we hypothesize that the critical threshold $p_c$ is set to be 11 values, i.e., 0.55, 0.56, $\ldots$, and 0.65. For each raw configuration, if the permeability is smaller than $p_c$, there will exist no percolating cluster and its label will be marked as 0; otherwise, there will exist at least one percolating cluster and its label will be marked as 1. 

To detect physic characteristics, we use the fourth kinds of CNNs (CNNs-\uppercase\expandafter{\romannumeral4}) with the structure in Fig.~\ref{fig:cnn} for 11 preset critical thresholds from 0.55 to 0.65. Note that the output layer use a sigmoid activation function expressed as $\bm{a}=1/(1+e^{-\bm{x}})$, to make sure the outputs are between 0 and 1. Another critical thing to pay attention is that the He normal distribution initializer \cite{HeNormal} and L2 regularization \cite{L2} are used in the layer of ``Conv1'', ``Conv2'', and ``FC'' on CNNs-\uppercase\expandafter{\romannumeral4}. To avoid overfiting, in addition to L2 regularization, we also use a dropout layer with a dropout rate of 0.5 on ``FC''. A Mini batch size of 512 and a learning rate of $10^{-4}$ are chosen while training CNNs-\uppercase\expandafter{\romannumeral4}. The binary cross-entropy (see Eq.~\ref{eq:BinaryCrossEntropy}) is taken as the loss function on CNNs-\uppercase\expandafter{\romannumeral4}. Another metric, used to measure the performance of CNNs-\uppercase\expandafter{\romannumeral4}, is the binary accuracy (see Eq.~\ref{eq:binary_accuracy}). 
The other hyper-parameters are the same as the CNNs-\uppercase\expandafter{\romannumeral1}, CNNs-\uppercase\expandafter{\romannumeral2}, and CNNs-\uppercase\expandafter{\romannumeral3}.
\begin{align}
\text{BinaryAccuracy}= \sum_{i=1}^{M}\frac{n_{(y_i^{\text{pred}}==y_i^{\text{raw}})}}{n_{y_i^{\text{pred}}}}
\label{eq:binary_accuracy}.
\end{align}

\begin{figure}[htbp]
  \noindent\includegraphics[width=0.48\textwidth]{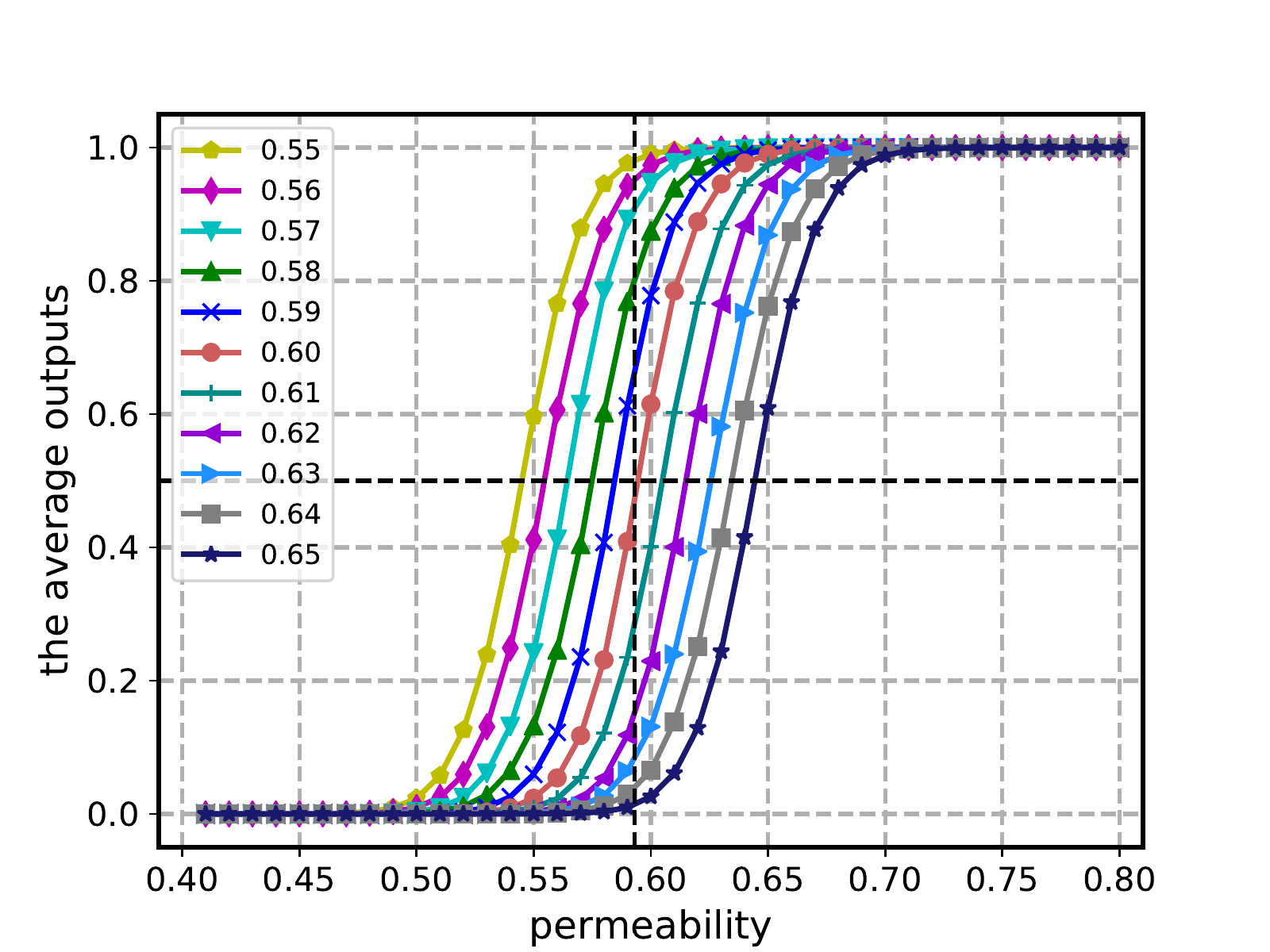}
  \caption{\label{fig:cnn-critical} The 11 curves show that the average outputs shifts when the preset threshold changes from 0.55 to 0.65. The average outputs and the threshold of phase transitions deduce from different preset threshold values on the percolation model by CNNs-\uppercase\expandafter{\romannumeral4}.}
\end{figure}

Turning now to the experimental evidence on the inference ability of capturing relevant physic features. After obtaining the well-trained CNNs-\uppercase\expandafter{\romannumeral4} with high accuracy (see Fig. S. 7), we obtain the outputs by pouring the raw 40,000 configurations $\{\bm{X}_1,\bm{X}_2,\ldots,\bm{X}_M\}$ into CNNs-\uppercase\expandafter{\romannumeral4}. The statistical average of the outputs is calculated according to the 40 independent permeability values $\{0.41,0.42,\ldots,0.80\}$. The results of the correlational analysis are shown in Fig.~\ref{fig:cnn-critical}. We set the horizontal dashed line as the threshold value 0.5. Hence, each curve is divided into two parts by the horizontal dashed line. The lower part indicates that the percolation system is not penetrated; while the upper part implies that the percolation system is penetrated. The crosspoint, where the horizontal dashed line and the red curve are intersected, has a permeability value of 0.594, which is very close to the theoretical value of 0.593 that is marked by the vertical dashed line in Fig.~\ref{fig:cnn-critical}. Remarkably, the critical transition point can be calculated by CNNs-\uppercase\expandafter{\romannumeral4} with the preset value of 0.60 for the sampling interval of 0.01. Therefore, CNNs-\uppercase\expandafter{\romannumeral4} with the preset threshold value of 0.60 is the most effective model. In further studies, the preset threshold value may need to be enhanced by smaller sampling intervals for higher precision.

\section{\label{sec:conclusion}Conclusions}

As machine learning approaches have become increasingly popular in phase transitions and critical phenomena, predecessors have pointed out that these approaches can capture physic characteristics. However, previous studies about identifying physical characteristics, especially order parameter and critical threshold, need to be further mutually validated. To highlight the possibility of effectiveness by machine learning methods, we conduct a much more comprehensive research than predecessors to reassess the machine learning approaches in phase transitions and critical phenomena. 

Our results show the effectiveness of machine learning approaches in phase transitions and critical phenomena than previous researchers. Precisely, we use CNNs-\uppercase\expandafter{\romannumeral1}, CNNs-\uppercase\expandafter{\romannumeral2} and CNNs-\uppercase\expandafter{\romannumeral3} to simulate the two order parameters, and the permeability values. To identify whether or not CNNs-\uppercase\expandafter{\romannumeral1} and CNNs-\uppercase\expandafter{\romannumeral2} are robust against noise, we add a proportion of the noises for the two order parameters. To validate the robustness of CNNs-\uppercase\expandafter{\romannumeral1}, CNNs-\uppercase\expandafter{\romannumeral2} and CNNs-\uppercase\expandafter{\romannumeral3}, we also use VAE and cVAE to generate new configurations that are slightly different from their raw configurations. After pouring the new configurations into the CNNs-\uppercase\expandafter{\romannumeral1}, CNNs-\uppercase\expandafter{\romannumeral2}, and CNNs-\uppercase\expandafter{\romannumeral3}, we achieve the results that these models are robust against noise. 

However, after we use PCA to reduce the dimension of the raw configurations and make a statistically significant linear correlation between the first principal component and the permeability values, no statistically significant linear correlations are found between the first principal component and the two order parameters. Clearly, the first principal component fails to be regarded as an order parameter in the two-dimensional percolation model. To identify order parameter, we use the fourth kinds of CNNs, i.e., CNNs-\uppercase\expandafter{\romannumeral4}. The results show that CNNs-\uppercase\expandafter{\romannumeral4} can identify new order parameter when the preset threshold value is 0.60. Surprisingly, we find that the critical transition point value is 0.594 by CNNs-\uppercase\expandafter{\romannumeral4}.

Although these machine learning methods are valid to explore the physical characteristics in the percolation model, the current study may still have some inevitable limitations that prevent us from making an overall judgement by these methods on the other models of phase transitions and critical phenomena. In other words, it must be acknowledged that this research is based on the two-dimensional percolation model.We are not sure of the usefulness of applying our methods to the other models. Consequently, our methods in this study may open an opportunity to other models on phase transitions and critical phenomena for further research.

\begin{acknowledgments}
The authors gratefully thank Yicun Guo for revising the manuscript. We also thank Jie-Ping Zheng and Li-Ying Yu for helpful discussions and comments. The work of S.Cheng, H.Zhang and Y.-L.Shi is supported by Joint Funds of the National Natural Science Foundation of China (U1839207). The work of F.He and K.-D.Zhu is supported by the National Natural Science Foundation of China (No.11274230 and No.11574206) and Natural Science Foundation of Shanghai (No.20ZR1429900).
\end{acknowledgments}


\nocite{*}

\bibliography{manuscript}

\end{document}